\documentclass[letterpaper,11pt]{article}

\usepackage{fancyhdr}
\usepackage[normalem]{ulem}
\usepackage{amsmath}
\usepackage{amsbsy}
\usepackage{amssymb}
\usepackage{amsthm}
\usepackage{amscd}
\usepackage{longtable}
\usepackage{amsfonts}
\usepackage{supertabular}
\usepackage{graphics}
\usepackage{graphicx}
\usepackage{verbatim}
\usepackage{caption}
\usepackage{subcaption}
\usepackage{epsfig}	
\usepackage{xspace}
\usepackage{euscript}
\usepackage{alltt}
\usepackage{boxedminipage}
\usepackage{morefloats}
\usepackage{times}
\usepackage[colorlinks]{hyperref}
\usepackage[authoryear,round]{natbib}
\usepackage{setspace}
\usepackage{authblk}
\usepackage{siunitx}
\usepackage{algorithm}
\usepackage{algorithmic}
\usepackage{relsize,exscale}

\hypersetup{colorlinks=true,linkcolor=black,citecolor=black}

\begin{document}

\title{Loading-unloading contact law for micro-crystalline cellulose particles under large deformations}
\author[1]{Yasasvi~Bommireddy}
\author[1]{Ankit~Agarwal}
\author[2]{Vikas~Yettella}
\author[2]{Vikas~Tomar}
\author[1,3]{Marcial~Gonzalez \thanks{Corresponding author at: School of Mechanical Engineering, Purdue University, West Lafayette, IN 47907, USA. Tel.:~+1~765 494 0904. Fax: +1 765 496 7537 \\ \indent E-mail address: marcial-gonzalez@purdue.edu (M. Gonzalez)}}
\affil[1]{\small School of Mechanical Engineering, Purdue University, West Lafayette, IN 47907, USA}
\affil[2]{\small School of Aeronautics and Astronautics, Purdue University, West Lafayette, IN 47907, USA}
\affil[3]{\small Ray W. Herrick Laboratories, Purdue University, West Lafayette, IN 47907, USA}

\maketitle

\begin{abstract}
A semi-empirical mechanistic contact law for micro-crystalline cellulose (Avicel PH-200) particles is proposed and characterized experimentally using force-displacement curves obtained from diametrical compression of single particles. The concepts of a shape factor and a master contact law are introduced first for elastic ellipsoidal particles, and subsequently generalized to plastic irregular particles. The proposed loading-unloading contact law is a function of three characteristic diameters (lengths of the principal axes of an approximated ellipsoid), a geometric parameter associated with the loading condition, three plastic and one elastic material properties. The force-displacement curves obtained using a micro-compression tester exhibit an apparent strain-hardening at distinctly different strain values, which is captured by the shape factor function and its geometric parameter. The three plastic material properties are log-normal distributions estimated from the loading experimental curves, while the elastic property is estimated from the unloading experimental curves. The study shows a very good agreement between predictions of the calibrated loading-unloading contact law and the experimental values.
\end{abstract}
\begin{center}
	\includegraphics[scale=0.61,angle=0]{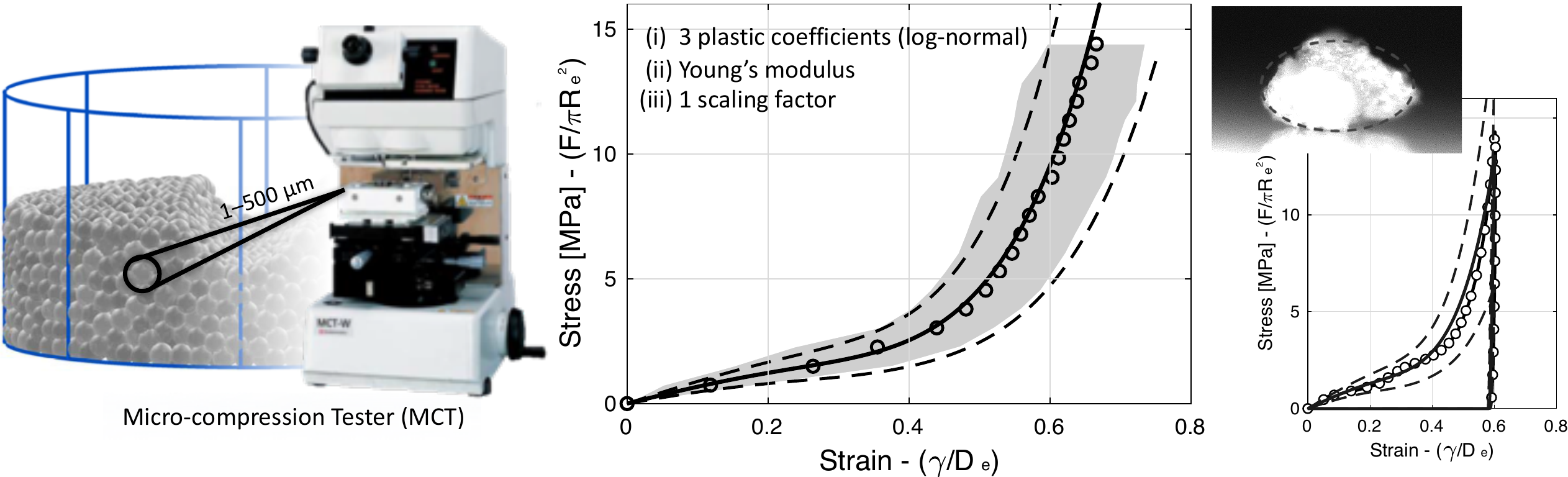}
\end{center}

\section{Introduction}

Over the last decade, pharmaceutical industry has undergone a paradigm shift \citep{ierapetritou_perspectives_2016} towards adoption of efficient process, control and material usage strategies in manufacturing, commonly referred to as Quality by Design (QbD) \citep{yu_understanding_2014,etzler_powder_2013}, which provide guidelines for optimal control over quality of the final product by acquiring a detailed understanding of the effect of process parameters on raw material attributes through (semi) empirical models. These strategies are further complemented by Quality by Control (QbC) \citep{su_QbC_2018} techniques, which are more prevalent in the continuous manufacturing practices as they provide increased flexibility to the design space and efficiency in characterization processes \citep{su_systematic_2017,SuBommireddy2018}.

Understanding the connection between mechanical behavior of a single particle and tabletability of the powder, i.e., the ability of the powder to gain strength under confined pressure, is essential for realizing QbD and QbC objectives \citep{simek_comparison_2017,Yi2018}. An alternative approach utilizing empirical models has been extensively attempted, albeit with limited success, to try to capture the stress-strain relationship of a powder bed under load. Heckel and Kawakita \citep{denny_compaction_2002,simek_comparison_2017} models are amongst the ones commonly used to relate effective compaction parameters with deformation mechanisms and material properties of single particles. In general, these empirical relationships cannot be interpreted unambiguously, unless a plethora of empirical compaction equations is selectively used \citep{nordstrom2012protocol}. Similar empirical efforts have been made, with rather limited success, to relate effective strength parameters of the compact with particle material properties, e.g., by means of the Leuenberger equation \citep{kuentz_new_2000}.

The challenges in achieving a systematic experimental characterization of micro-sized particles under confined conditions and large deformations are associated to the small size and irregular morphology of the particles, typical of pharmaceutical powders. Diametrical compression of single micro-particles using indentation equipment or Micro-Compression Testers (MCT) is one of the most common confinement techniques used to characterize the mechanical properties of various materials, especially polymers \citep{Egholm-2006,He-2009,Tanaka2007,Liu-1998}. \citet{Tanaka2007a} used a micro compression equipment to study the compaction behavior of individual 50 micron alumina granules with PVA (Poly Vinyl Alcohol) or PAA (Poly Acrylic Acid) as the binder matrix. In the context of pharmaceutical materials, \citet{Yap2008} studied the diametrical compression of single particles of various pharmaceutical excipients to relate the mechanical properties characterized at particle-scale to bulk compression properties of the materials. The technique has also been used to characterize visco-elastic properties of agarose \citep{Yan2009} and alginate \citep{Nguyen2009,Wang2005} microspheres. A major recent development in confinement methods for single particles has been made by \citet{Jonsson-2015}, who have developed a novel apparatus for triaxial testing of single particles, providing a more realistic insight into the behavior of individual particles under confinement. The apparatus has been used to investigate and characterize the yield and fracture behavior of micro-crystalline cellulose (MCC) granules \citep{Jonsson-2016}, highlighting the critical differences in their fracture behavior during triaxial and uniaxial compression, and correlating the triaxial response of single particles with bulk powder compression response.

In most of the mentioned works, the shape of studied particles is either spherical or assumed to be spherical, primarily for employment of existing analytical contact formulations that were developed specifically for spherical particles due to obvious geometrical simplicity. The elastic mechanical properties, typically the Young's Modulus and Poisson's ratio, are commonly characterized using Hertz contact law \citep{Hertz-1881}, while the inelastic properties, typically the hardness or representative strength of the material, are characterized using various contact models developed for describing spherical indentation \citep{Tabor-1951, Johnson1985,Biwa-1995} and contact of inelastic solids of revolution \citep{Storakers-1997}. It is important to note that these contact models are applicable only in the range of small deformations due to various limiting assumptions, such as independent contacts and simplified contact surface profile curvatures \citep{Gonzalez2012,AGARWAL201826}. 

In this work, we restrict attention to micro-crystalline cellulose (Avicel PH-200) particles, with average diameters ranging between $50$ and $300~\mu$m, under large diametrical compression. Micro-crystalline cellulose is one of the widely used binding materials in pharmaceutical powder blends for die-compaction of solid tablets \citep{jivraj_overview_2000}. MCC particles exhibit significant plastic deformations under diametrical compression and elastic relaxation during unloading, which makes them relevant for this study \citep{mashadi_characterization_1987,mohammed_study_2006}. We address the first set of challenges described above by using a Shimadzu MCT-510 micro-compression tester equipped with a flat punch tip of $500~\mu$m in diameter, a top microscope, and a side camera capable of recording the deformation process. The micro-compression tester is capable of applying loading-unloading cycles under load control mode within a wide loading range of $9.8$ to $4803$~mN and a displacement range of $0$ to $100~\mu$m, at a minimum increment of $0.001~\mu$m. We then characterize particle size and shape using the top microscope and the side camera, and record accurate force-displacement curves over a range of large deformations. The characterization of these irregular particles under lateral confinement clearly increases the experimental complexity and, even though it provides complementary information for building a semi-empirical mechanistic model, it will be beyond the scope of this work.

Despite the challenges described above, it is expected, as is the case for a wide range of physical responses, that a special structure exists in the response of micro-sized particles under large diametrical compression, which can be discovered and exploited to create a semi-empirical mechanistic model. Specifically, we assume that the force-displacement response exhibits a special structure known as an active subspace (see \citet{tripathy2016gaussian} and references therein), i.e., a manifold of the stochastic space of inputs (such as particle size, shape, surface roughness, internal porosity, material properties, loading conditions and confinement) characterized by maximal variation of the multivariate response function (i.e., the contact force). We identify this low dimensional manifold based on mechanistic understanding of the problem, project onto it the high-dimensional space of inputs, and link the projection to the measured output contact force to build a semi-empirical mechanistic contact law. It is worth noting that the high-dimensional space of inputs is typically not amenable to a full experimental characterization and thus some components of the active subspace need to be estimated as part of the process of building the response function. Specifically, we estimate three plastic and one elastic material properties, and one geometric parameter associated with the loading condition through the shape factor function. Finally, we account for model inaccuracies and/or imperfections by endowing the three plastic material properties with a probability distribution, namely a log-normal distribution.

It bears emphasis that a semi-empirical mechanistic contact law for elasto-plastic particles is also relevant to three-dimensional particle mechanics calculations \citep{Gonzalez2012,gonzalez2016microstructure,yohannes2016evolution,yohannes2017discrete,gonzalez2018statistical}. The particle mechanics approach enabled the prediction of microstructure evolution during the three most important steps of powder die-compaction (namely compaction, unloading, and ejection) using generalized loading-unloading contact laws for elasto-plastic spheres with bonding strength \citep{Gonzalez2018generalized}. Moreover, these detailed calculations enabled the development of microstructure-mediated process-structure-property-performance interrelationships for QbD and QbC product development and process control, which depend on a small number of parameters with well-defined physical meanings. Therefore, the work presented in this paper, in combination with particle mechanics calculations, can contribute towards establishing the relationship between particle-level material properties and tablet performance.

The paper is organized as follows. Contact mechanics formulations for elastic ellipsoidal particles are reviewed in Section~\ref{Section-EllipsoidalParticles}, followed by the introduction of shape factor and master contact law concepts, which are subsequently generalized to plastic irregular particles. The experimental analysis of MCC particles under diametrical compression is discussed in Section~\ref{Section-DiamtricalCompression}. Plastic shape factor for MCC particles is proposed in Section~\ref{Section-ShapeFactor}, followed by the description of an optimization procedure used to determine the geometric parameter associated with the loading condition. in Section~\ref{Section-MasterLaw}, a master contact law for micro-crystalline cellulose particles is proposed, and log-normal distributions for the three plastic material properties are estimated from experimental data. Section~\ref{Section-Loading-Unloading} extends the work to loading-unloading contact laws, and Section~\ref{Section-Results} shows the comparison between predictions of the calibrated contact law and the experimental values. Finally, a summary is presented in Section~\ref{Section-Summary}.

\section{Contact mechanics formulations for ellipsoidal particles}
\label{Section-EllipsoidalParticles}

\begin{figure}[t]
	\centering 
	\includegraphics[scale=0.7]{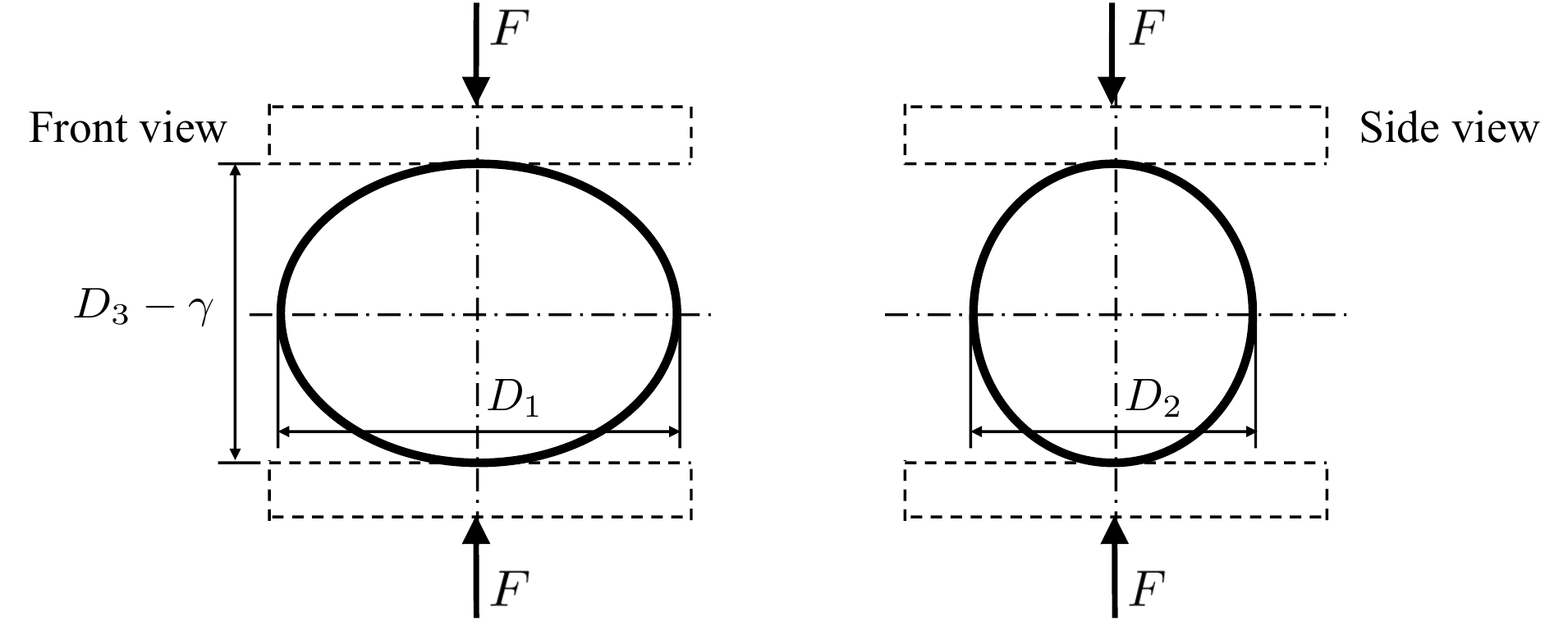}
	\caption{Front and side view of an ellipsoidal particle under diametrical compression in the direction of $D_3$.}
	\label{Fig-ParticleDimensions}
\end{figure}

The contact mechanics of ellipsoidal elastic particles, characterized by three diameters $D_1$, $D_2$ and $D_3$ (Fig.~\ref{Fig-ParticleDimensions}), indicates that the contact force $F$ applied on a particle by two rigid plates in the direction of $D_3$ is given by \citep{zheng_contact_2013,Johnson1985}

\begin{equation}
F 
= 
\frac{4 E \sqrt{R_e}}{3(1-\nu^2)}
\left[\eta_e(D_1,D_2)~\gamma \right]^{3/2}
\label{Eqn-EllipsoidalHertz}
\end{equation}
where $\gamma$ is the relative displacement of the plates, $E$ and $\nu$ are the Young's modulus and the Poisson's ratio respectivley, $R_e=D_1 D_2/2D_3$ is the effective radius of the ellipsoidal particle, and $\eta_e$ is a shape factor dependent on $D_1$ and $D_2$ , given by
\begin{equation}
\frac{1}{\eta_e(D_1,D_2)}
=
1-\left[\left(\frac{D_1}{D_2}\right)^{0.1368}-1\right]^{1.531}
\end{equation}
By defining effective stress and strain as
\begin{equation}
\sigma = \frac{F}{\pi R_e^2}
\label{Eqn-Stress}
\end{equation}
\begin{equation}
\epsilon = \frac{\gamma}{2 R_e}
\label{Eqn-Strain}
\end{equation}
the following stress-strain relationship $\sigma(\eta_e \epsilon)$ is recovered
\begin{equation}
\sigma 
:=
\sigma(\eta_e \epsilon)
=
\frac{8\sqrt{2} E}{3\pi (1-\nu^2)}
\left[\eta_e(D_1,D_2)~\epsilon \right]^{3/2}
\label{Eqn-EllipsoidalHertz2}
\end{equation}
which depends on material properties ($E$ and $\nu$) and particle geometry ($D_1$ and $D_2$). It is worth noting that for spherical particles, the effective radius $R_e$ is equal to the radius of the sphere, the elastic shape factor $\eta_e$ simplifies to 1, and Hertz theory is recovered. Another noteworthy consideration from the above equation is that $\eta_e$ is independent of the particle height $D_3$. This is an artefact of the \emph{local character} of the formulation, which assumes that the contacts evolve locally and independently without any interactions with other contacts on the same particle. Consequently, evolution of the contact is dependent only on the in-plane dimensions, i.e. $D_1$ and $D_2$. In the context of confined granular systems, this assumption has been found to hold only during the inital stages of compression \citep{Mesarovic-2000}. For linear-elastic spherical particles, the nonlocal contact formulation by \citet{Gonzalez2012}, and its later extension by \citet{AGARWAL201826} to correct particle contact areas, is a significant contribution towards relaxation of this limiting assumption. The formulation invokes the principle of superposition to describe the normal contact deformation and evolution of inter-particle contact area as a sum of local (i.e., Hertzian) deformation and nonlocal deformations generated by other contact forces acting on the same particle. For elasto-plastic materials, development of a closed-form analytical formulation for general contact configurations that can describe contact interactions at large deformations \citep{Tsigginos2015} still remains an open problem. Initial progress in this regard has been made by \citet{Frenning2013}, who proposed a truncated sphere model applicable for small to moderate deformations, that utilizes the plastic incompressibility assumption to relate the average pressure in the particle due to elastic volumetric strain to the mean pressure generated at the particle contacts.

\begin{figure}[b!]
	\centering
	\begin{subfigure}[b]{0.4\textwidth}
		\includegraphics[width=\linewidth,trim=60 70 60 119, clip]{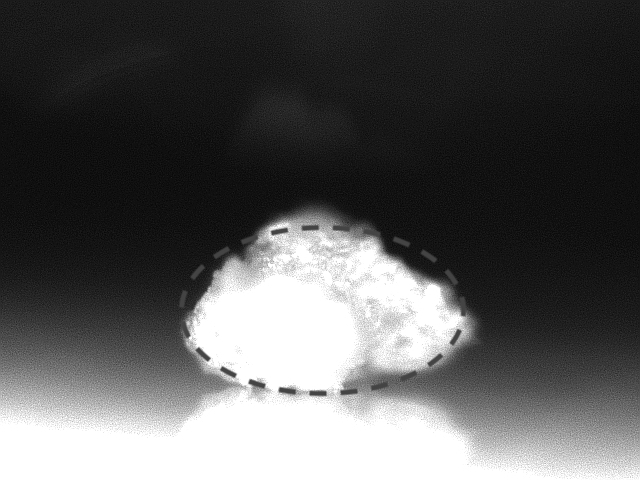}	
	\end{subfigure}
	\hspace{30pt}
	\begin{subfigure}[b]{0.4\textwidth}
		\includegraphics[width=\linewidth,trim=10 0 10 130, clip]{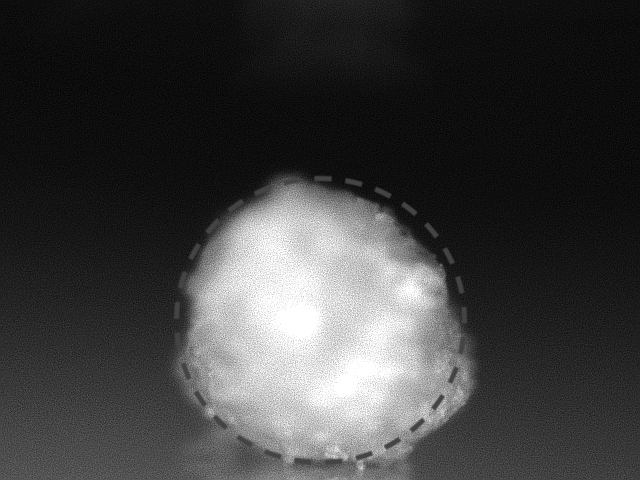}
	\end{subfigure}
	\caption{Side view of two Avicel PH-200 irregular particles, cf. Fig.~\ref{Fig-ParticleDimensions}. Left: $D_1=155~\mu$m, $D_2=217~\mu$m, $D_3=136~\mu$m. Right: $D_1=213~\mu$m, $D_2=196~\mu$m, $D_3=186~\mu$m.}
	\label{Fig-SideViewExamples}
\end{figure}

In this work, we interpret the function $\eta_e$ as the elastic shape factor that depends solely on particle geometry and loading condition, and the function $\sigma(\cdot)$ as the master contact behavior of the particle that depends solely on the particle deformation behavior (i.e., on its elastic behavior). Furthermore, we generalize this interpretation to plastic ellipsoidal particles as follows
\begin{equation}
\sigma := \sigma(\eta_p \epsilon)
\label{Eqn-MasterPlastic-1}
\end{equation}
where the plastic master behavior depends on the deformation mechanisms of plastic particles and the shape factor is 
\begin{equation}
\eta_p := \eta_p(D_1/D_3, D_2/D_3) \mbox{~~such that~~}\eta_p(1,1)=1
\label{Eqn-ShapeFactor-1}	
\end{equation}
It is worth noting that the plastic shape factor depends on $D_3$ in addition to $D_1$ and $D_2$, which emphasizes its applicability to large deformations. In order to demonstrate this generalization, we develop an experimental and numerical procedure to identify (i) the plastic master contact behavior $\sigma(\cdot)$ for micro-crystalline cellulose (Avicel PH-200) under large plastic deformations, and (ii) the shape factor $\eta_p$ for irregular particles that can be approximated by ellipsoids (Fig.~\ref{Fig-SideViewExamples}). 

It is important to understand the characteristics of micro-crystalline cellulose particles and their dependency on environmental and process variables (see, e.g.,  \citet{sun_mechanism_2008,inghelbrecht_roller_1998,westermarck_microcrystalline_1999}). These properties gather further importance in implementing QbD and QbC strategies, which require the understanding of raw material attributes for robust in-line monitoring and control of manufacturing process \citep{thoorens_microcrystalline_2014}. Avicel has excellent compactibility,  high dilution potential and superior disintegration properties, making it a prime filler material for the challenging direct compression tabletting for a wide range of active pharmaceutical ingredients (APIs)  \citep{jivraj_overview_2000,celik_compaction_1996,doelker_comparative_1993,reier_microcrystalline_1966}. Different grades of micro-crystalline cellulose exist in the pharmaceutical industry, increasing the flexibility of its usage to not only direct compression but also to dry granulation and wet granulation processes \citep{inghelbrecht_roller_1998,westermarck_microcrystalline_1999}. Some of the grades have different functional usages and not just physical variations such as to counter the weakening effect of lubricants on tablet strength \citep{van_veen_compaction_2005}.

\section{Diametrical compression of micro-crystalline cellulose particles}
\label{Section-DiamtricalCompression}
\begin{figure}[t]
	\centering
	\includegraphics[width=0.27\textwidth]{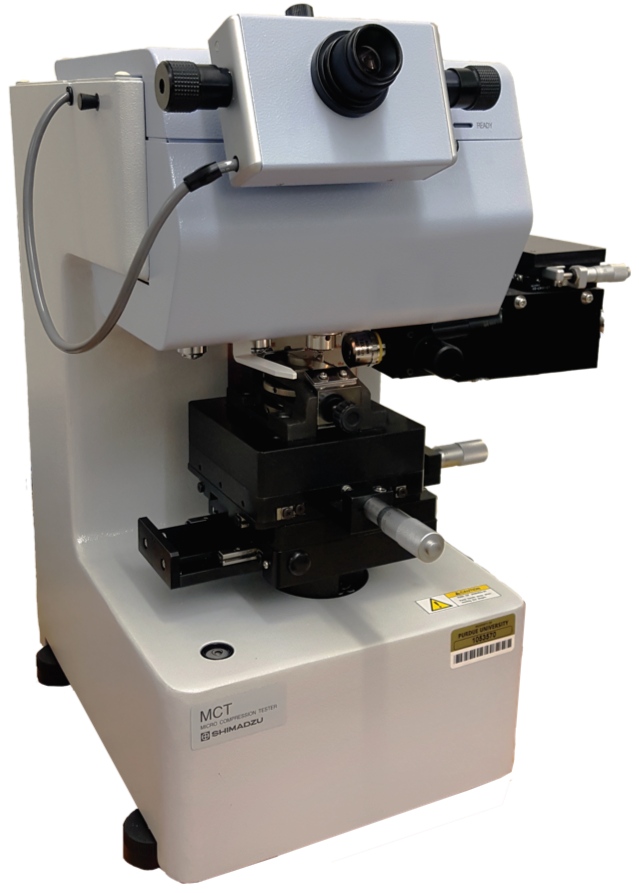}
	\caption{Shimadzu MCT-510 micro-compression tester.}
	\label{Fig-MCT}
\end{figure}

\begin{figure}[b!]
	\centering 
	\includegraphics[width=0.55\textwidth]{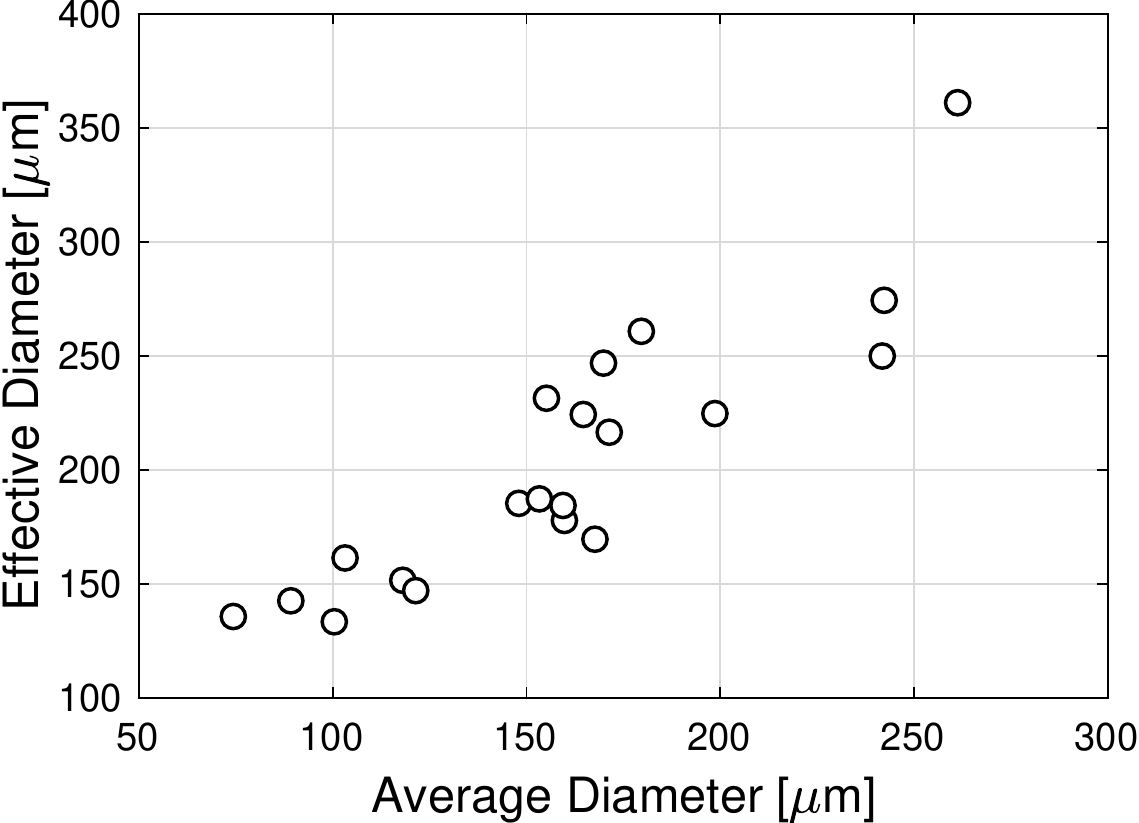}
	\caption{Dimensions of tested Avicel PH-200 particles. The ratio between the average diameter, i.e., $(D_1+D_2+D_3)/3$, and effective diameter, i.e., $D_1 D_2/D_3$, appears to be constant and about 1.5.}
	\label{Fig-DimensionStatistics}
\end{figure}

Micro-crystalline cellulose (Avicel PH-200) particles were tester under diametrical compression using a Shimadzu MCT-510 micro-compression tester (see Fig.~\ref{Fig-MCT}) equipped with a flat punch tip of $500~\mu$m in diameter. Load-unload tests were performed on 20 particles at a constant stress rate of $2$~mN/s, with maximum applied stress in the range of $7-16$~MPa. The characteristic diameters $D_1$ and $D_2$ were measured from the top view using the in-built microscope, while $D_3$ was calculated from a side view obtained using the side camera. The obtained dimensions for all particles are listed in Table \ref{Table-MCC-Dimensions}, while morphology of the tested particles is illustrated in Fig.~\ref{Fig-DimensionStatistics}, showing the relationship between their average diameter, i.e., $(D_1+D_2+D_3)/3$, and their effective diameter, i.e., $D_1 D_2/D_3$. It is interesting to note that this relationship has a linear trend, and that the ratio of the two values is about 1.5 for Avicel PH-200. 

\begin{table}[t]
	\centering
	\begin{tabular}{|c|c|c|c|}
		\hline
		Particle No. &  $D_1 (\mu\mathrm{m})$  &  $D_2 (\mu\mathrm{m})$ &  $D_3 (\mu\mathrm{m})$ \\
		\hline
		$1$ & $213.54$ & $196.07$ & $186.33$ \\
		$2$ & $155.32$ & $217.51$ & $136.87$ \\
		$3$ & $168.74$ & $171.64$ & $125.15$ \\
		$4$ & $97.56$ & $100.98$ & $69.09$ \\
		$5$ & $238.47$ & $249.24$ & $237.81$ \\
		$6$ & $93.8$ & $76.46$ & $52.85$ \\
		$7$ & $160.44$ & $167.7$ & $151.3$ \\
		$8$ & $114.1$ & $114.54$ & $80.98$ \\
		$9$ & $168.31$ & $144.54$ & $131.27$ \\
		$10$ & $133.8$ & $117.12$ & $103.38$ \\
		$11$ & $122.69$ & $131.73$ & $109.91$ \\
		$12$ & $121.04$ & $205.93$ & $133.08$ \\
		$13$ & $185.1$ & $206.98$ & $146.91$ \\
		$14$ & $180.35$ & $173.88$ & $139.81$ \\
		$15$ & $173.26$ & $157.3$ & $147.79$ \\
		$16$ & $296.34$ & $267.76$ & $219.78$ \\
		$17$ & $219.5$ & $281.94$ & $225.55$ \\
		$18$ & $194.97$ & $167.95$ & $151.23$ \\
		$19$ & $105.57$ & $109.25$ & $86.46$ \\
		$20$ & $154.12$ & $182.82$ & $166.13$ \\
		\hline
	\end{tabular}
	\caption{Obtained characteristic dimensions of the studied Avicel PH-200 particles}
	\label{Table-MCC-Dimensions}
\end{table}

\begin{figure}[b!]
	\centering
	\begin{subfigure}[b]{0.3\textwidth}
		\includegraphics[width=\linewidth,trim=100 65 100 135, clip]{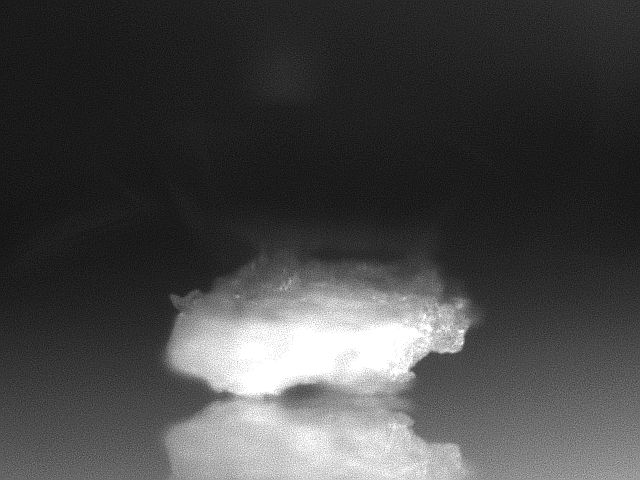}	
	\end{subfigure}
	\begin{subfigure}[b]{0.3\textwidth}
		\includegraphics[width=\linewidth,trim=100 50 100 150, clip]{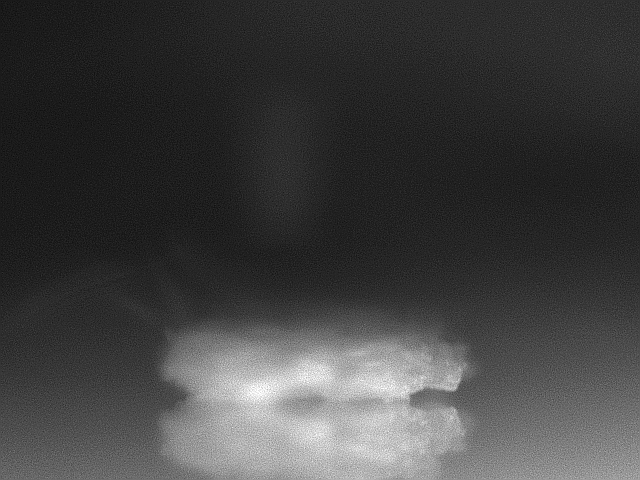}
	\end{subfigure}
	\begin{subfigure}[b]{0.3\textwidth}
		\includegraphics[width=\linewidth,trim=100 50 100 150, clip]{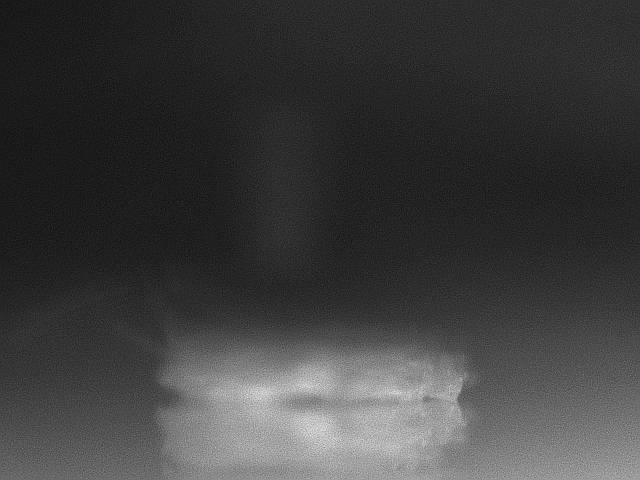}
	\end{subfigure}
	\caption{Avicel PH-200 particle at three increasingly higher levels of diametrical compression (from left to right).}
	\label{Fig-DeformationSideVIew}
\end{figure}
\begin{figure*}[t]
	\centering
	\includegraphics[width=0.55\textwidth]{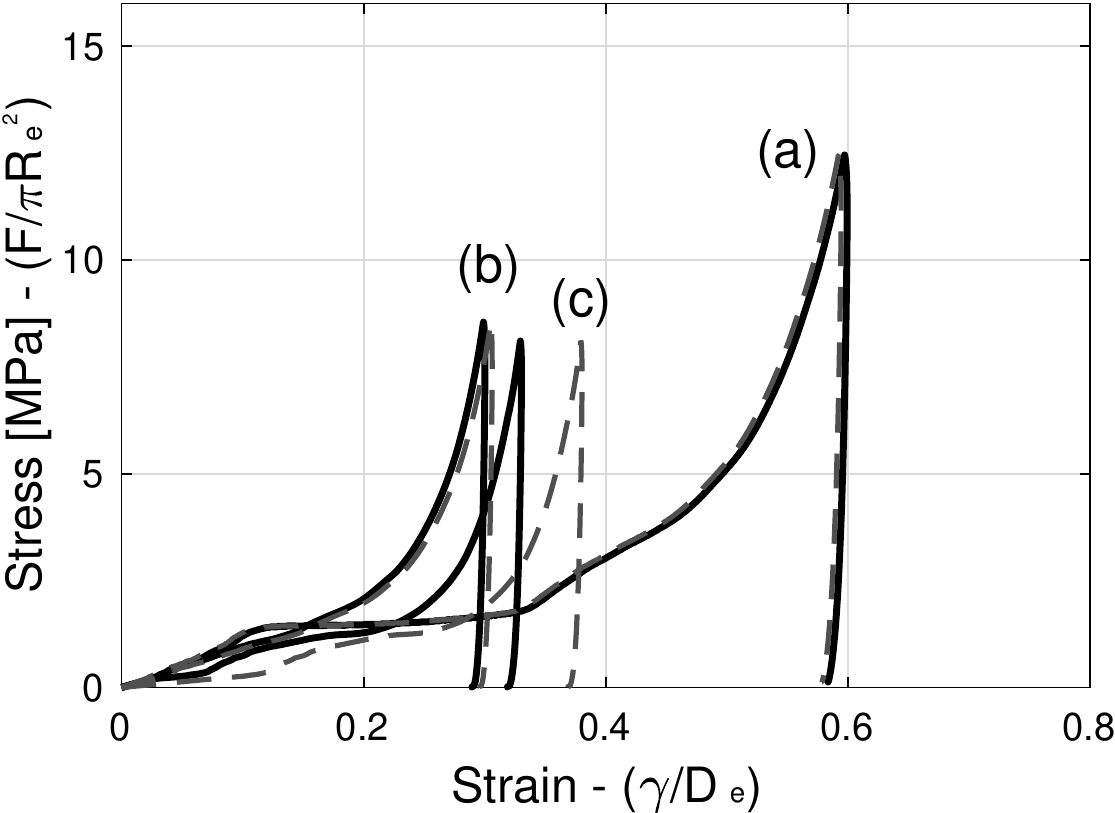}
	\caption{Diametrical loading-unloading curves for particles 1, 2 and 3. Solid lines correspond to the raw measurements of force $F$ and displacement $\gamma$ normalized by the corresponding effective diameter $D_e$ to obtain stress and strain, respectively. The dashed lines correspond to the adjusted loading-unloading curves, using a stress threshold of $0.2$~MPa.}
	\label{Fig-ExampleCurves}
\end{figure*}

As an illustrative example, Fig.~\ref{Fig-DeformationSideVIew} shows an Avicel PH-200 particle at three increasingly higher levels of diametrical compression. It is evident from the figure that the particle undergoes large deformations without apparent brittle failure, but rather permanent plastic deformation, as confirmed from the unloading elastic curves in Fig.~\ref{Fig-ExampleCurves}. It is also evident from the figures that surface irregularities are eliminated at low levels of diametrical compression. However, Fig.~\ref{Fig-ExampleCurves} illustrates that the strain required to eliminate these irregularities, which is measured from the first contact between the particle surface and the flat punch tip, is different for each particle and, therefore, is not predictable by a deterministic contact mechanics formulation such as equations~\eqref{Eqn-MasterPlastic-1}-\eqref{Eqn-ShapeFactor-1}. Therefore, in this study, we adopted a threshold stress of $0.2$~MPa to separate stochastic from deterministic deformation behaviors (i.e., to separate the initial deformation behavior dominated by uncharacterized surface irregularities from the subsequent behavior dominated by plastic deformations of an equivalent ellipsoidal particle). Specifically, we assumed a linear stress-strain response for the initial deformation behavior, with strain at $0.2$~MPa equal to the averaged value obtained from the 20 diametrical compression tests. The stress-strain response of each tested particle was then adjusted by offsetting the strain such that all deformation curves had the same averaged strain value at $0.2$~MPa. Fig.~\ref{Fig-ExampleCurves} shows raw and adjusted curves for three different particles. It is worth noting that the stress of $0.2$~MPa corresponds to 1-3$\%$ of the maximum stress applied during the tests. The adjusted diametrical compression curves were subsequently used to determine the shape factor $\eta_p(D_1/D_3,D_2/D_3)$ and the master contact law $\sigma(\cdot)$ for micro-crystalline cellulose particles. 

\section{Plastic shape factor for micro-crystalline cellulose particles}
\label{Section-ShapeFactor}

We proposed in Section~\ref{Section-DiamtricalCompression} that the plastic shape factor for ellipsoidal plastic particles $\eta_p$ must depend solely on $D_1/D_3$ and $D_2/D_3$ and that it must simplify to 1 for spherical particles (Eqn.~\eqref{Eqn-ShapeFactor-1}). In addition, using Eqn.~\eqref{Eqn-MasterPlastic-1}, we propose that the stress-strain relationship $\sigma_i(\epsilon)$ of an ellipsoidal particle $i$ is related to the master contact law $\sigma(\epsilon)$ by
\begin{equation}
\sigma_i(\epsilon) 
=
\sigma( \eta_{p,i} \epsilon)
\implies
\sigma(\epsilon) 
=
\sigma_i(\epsilon/\eta_{p,i})
\label{Eqn-Master-StressStrain}
\end{equation}
Equivalently, the strain-stress relationship $\epsilon(\sigma)$ of the master contact law and that of an ellipsoidal particle $i$, $\epsilon_i(\sigma)$, are related by
\begin{equation}
\epsilon(\sigma) = \eta_{p,i} \epsilon_i(\sigma)
\label{Eqn-Master-StrainStress}
\end{equation}
Even though Eqns.~\eqref{Eqn-Master-StressStrain} and \eqref{Eqn-Master-StrainStress} are equivalent, we shall use the experimentally characterized curves $\epsilon_i(\sigma)$ to determine the master contact law $\epsilon(\sigma)$ and the plastic shape factor $\eta_p$. This choice is motivated by the observation that micro-crystalline cellulose particles exhibit an apparent strain-hardening at distinctly different strain values (see Fig.~\ref{Fig-ExampleCurves}). Hence, the domain of the function $\epsilon_i(\sigma)$, that is $[0, \sigma^\mathrm{m}_i]$, is quite similar for all the characterized particles---with $\sigma^\mathrm{m}_i$ being the maximum applied stress during diametrical compression of particle $i$. Moreover, Avicel PH-200 particles are inhomogeneous and not perfectly ellipsoidal; thus, a unique function \eqref{Eqn-Master-StrainStress} that holds true for all tested particles does not exist. Therefore, we propose that the master contact law is given by a normal distribution
\begin{equation}
\epsilon(\sigma) \sim \mathcal{N}(\bar{\epsilon}(\sigma),s^2_\epsilon(\sigma))
\end{equation}
with expectation given by the sample mean $\bar{\epsilon}$ and variance given by square of the sample standard deviation $s^2_\epsilon$, i.e.
\begin{eqnarray}
\bar{\epsilon}(\sigma)
&:=&
\frac{1}{\#\mathcal{S}_{\sigma}}\sum_{i\in\mathcal{S}_{\sigma}} \eta_{p,i} \epsilon_i(\sigma)
\label{Eqn-MasterMean}
\\
s_\epsilon^2(\sigma)
&:=&
\frac{1}{\#\mathcal{S}_{\sigma}-1}\sum_{i\in\mathcal{S}_{\sigma}} \left[ \eta_{p,i} \epsilon_i(\sigma) - \bar{\epsilon}(\sigma) \right]^2
\label{Eqn-MasterStd}
\end{eqnarray}
for all $\sigma$ such that $\#\mathcal{S}_{\sigma} > N/5$. In the above equations $N$ is the total number of tested particles (i.e., $20$ in this study), $\mathcal{S}_{\sigma}$ is the set of experiments for which a stress of $\sigma$ has been measured or reached before unloading, and $\#\mathcal{S}_{\sigma}$ is the cardinality or the number of elements of $\mathcal{S}_{\sigma}$. In order to achieve statistical significance, a sufficient number of particles needs to be tested. For the purpose of demonstrating the proposed method, we adopt such number to be $N/5$.
Next, we propose the following shape factor for micro-crystalline cellulose particles 
\begin{equation}
\eta_p(D_1/D_3,D_2/D_3) 
:=	
1 + a \left( 1 - \frac{D_3^2}{D_1 D_2} \right)
\label{Eqn-ShapeFactor}
\end{equation}
where $a$ is a geometric parameter associated with the loading condition, to be determined from the experimental strain-stress curves $\epsilon_i(\sigma)$. Specifically, the parameter $a$ is such that the relative standard deviation of $\epsilon$ is minimized over a range of stresses $[0,\sigma^\mathrm{m}_{\sigma,i}]$, i.e., $a$ minimizes
\begin{equation}
{\mathlarger{\sum}}_{i=1}^{N}
\frac{{\mathop{\mathlarger{\mathlarger{\int}}}_{0}^{\sigma^\mathrm{m}_{\sigma,i}}
		\left[\frac{1}{\#\mathcal{S}_{\sigma}}\sum_{j\in\mathcal{S}_{\sigma}} \eta_{p,j} \epsilon_j(\sigma) - {\eta_{p,i}} \epsilon_i(\sigma) \right]^2\mathrm{d}\sigma}}
{\mathop{\mathlarger{\int}}_{0}^{\sigma^\mathrm{m}_{\sigma,i}}
	\left[\frac{1}{\#\mathcal{S}_{\sigma}}\sum_{j\in\mathcal{S}_{\sigma}} \eta_{p,j} \epsilon_j(\sigma)\right]^2\mathrm{d}\sigma}
\label{Fig-DetermineCoeff-1}
\end{equation}
with $\sigma^\mathrm{m}_{\sigma,i}=\min\{\sigma^\mathrm{m}_i,\sigma^\mathrm{m}\}$, $\sigma^\mathrm{m}$ being the maximum stress value in the master contact law $\bar{\epsilon}(\sigma)$, i.e., in Eqn.~\eqref{Eqn-MasterMean}. In order to simplify the optimization process, we first identify a particle with $D_3^2/D_1 D_2$ close to 1 which is diametrically compressed at a high maximum stress and label it as particle $N$, and then rewrite Eqn.~\eqref{Fig-DetermineCoeff-1} using
$$
\frac{1}{\#\mathcal{S}_{\sigma}}\sum_{i\in\mathcal{S}_{\sigma}} \eta_{p,i} \epsilon_i(\sigma)
\approx
\eta_{p,N} \epsilon_N(\sigma)
$$
which leads to the following approximate minimization problem
\begin{equation}
a := \arg \min_{a} {\mathlarger{\sum}}_{i=1}^{N-1}
\frac{{\mathop{\mathlarger{\mathlarger{\int}}}_{0}^{\sigma^\mathrm{m}_{\sigma,i}}
		\left[\epsilon_N(\sigma) - \dfrac{\eta_{p,i}}{\eta_{p,N}} \epsilon_i(\sigma) \right]^2\mathrm{d}\sigma}}
{\mathop{\mathlarger{\int}}_{0}^{\sigma^\mathrm{m}_{\sigma,i}}
	\left[{\epsilon_N(\sigma)}\right]^2\mathrm{d}\sigma}
\label{Eqn-Minimization}
\end{equation}
with $\sigma^\mathrm{m}_{\sigma,i}=\min\{\sigma^\mathrm{m}_i,\sigma_N^\mathrm{m}\}$.  It is worth noting that the experimental curves are linearly interpolated and numerically integrated for the purpose of solving Eqn.~\eqref{Eqn-Minimization}.

The proposed optimization process results in a shape factor, Eqn.~\eqref{Eqn-ShapeFactor}, for the tested Avicel PH-200 particles with an optimal coefficient $a=1.43$. The master contact behavior for these particles is determined in the next section.

\section{Master contact law for micro-crystalline cellulose particles}
\label{Section-MasterLaw}

\begin{figure}[t]
	\centering
	\includegraphics[width=0.55\textwidth]{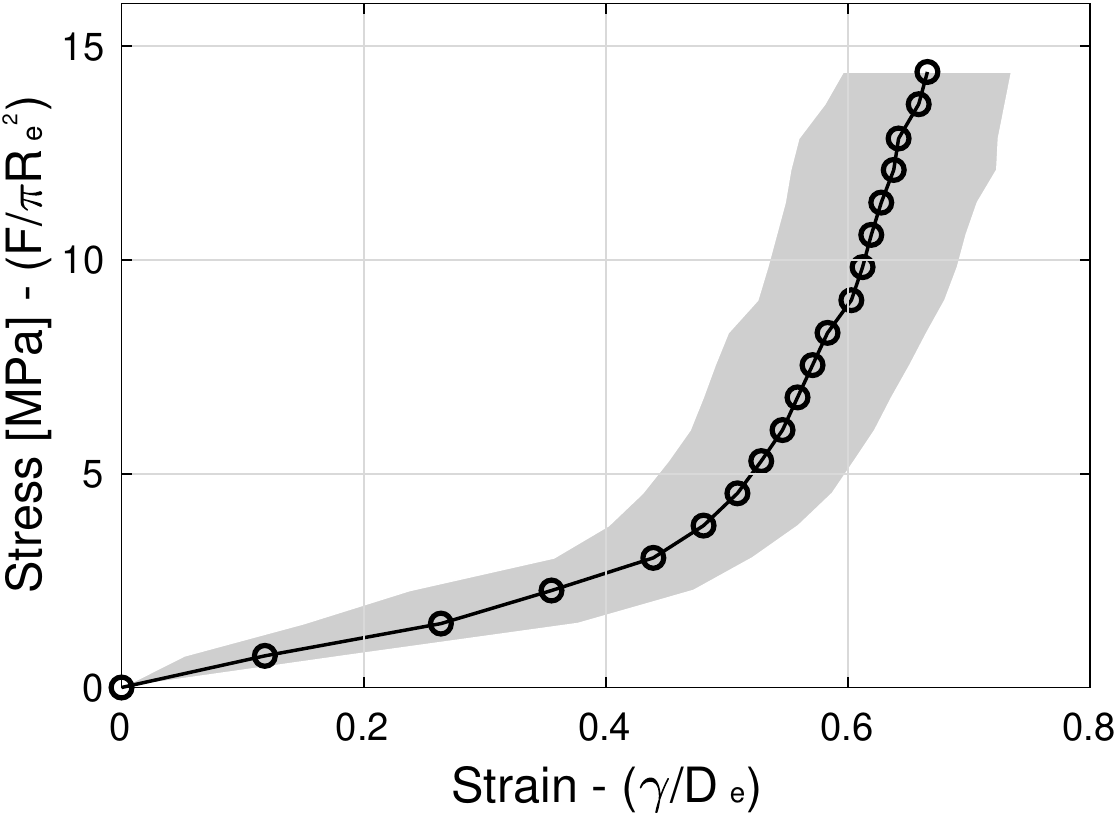}
	\caption{Master contact law $\epsilon \sim \mathcal{N}(\bar{\epsilon},s^2_\epsilon)$, for micro-crystalline cellulose particles. Symbols correspond to the mean. Shaded area corresponds to one standard deviation from the mean. The shape factor $\eta_p$ is given by Eqn.~\eqref{Eqn-ShapeFactor} with $a=1.43$.}
	\label{Fig-MasterPlot-1}
\end{figure}

We proposed in the previous section that the master contact law is given by a normal distribution $\epsilon(\sigma) \sim \mathcal{N}(\bar{\epsilon}(\sigma),s^2_\epsilon(\sigma))$, where the sample mean $\bar{\epsilon}(\sigma)$ and the sample standard deviation squared $s_\epsilon^2(\sigma)$ are given by Eqns.~\eqref{Eqn-MasterMean}-\eqref{Eqn-MasterStd} and the shape factor $\eta_p$ is given by Eqn.~\eqref{Eqn-ShapeFactor} with $a=1.43$. Therefore, the master contact law is readily available from the experimental curves and is shown in Fig.~\ref{Fig-MasterPlot-1}.

\begin{table}[b!]
	\centering
	\begin{tabular}{|c|c|c|}
		\hline
		&  $\mu_{C_i}$  &  $\sigma_{C_i}$ \\
		\hline
		$\log(C_1)$ &   $1.9027$ & $0.2985$ \\
		$\log(C_3)$ &  $1.1015$ & $4.8914 \times 10^{-9}$ \\
		$\log(C_5)$ & $0.9784$ & $0.0797$ \\
		\hline
	\end{tabular}
	\caption{Plastic coefficients $\log(C_i) \sim \mathcal{N}(\mu_{C_i},\sigma^2_{C_i})$ of the stress-strain relationship of the master contact law for micro-crystalline cellulose, Eqn.~\eqref{Eqn-Master-StressStrain2}.}
	\label{Table-StressStrain-Param}
\end{table}

The  stress-strain relationship $\sigma(\epsilon)$ of the master contact law for micro-crystalline cellulose particles is given by the inverse of $\epsilon(\sigma)$. We propose to approximate $\sigma(\epsilon)$ by
\begin{equation}
\sigma(\epsilon) / \mathrm{MPa} = C_1~\eta_p \epsilon - \left( C_3~\eta_p \epsilon \right)^3 + \left( C_5~\eta_p \epsilon \right)^5
\label{Eqn-Master-StressStrain2}
\end{equation}
with $\log(C_i) \sim \mathcal{N}(\mu_{C_i},\sigma^2_{C_i})$. Therefore, the expected value of $\sigma(\epsilon)$ is given by
\begin{equation}
\mathrm{E}\langle\sigma(\epsilon)\rangle / \mathrm{MPa} 
= 
\mathrm{e}^{\mu_{C_1}+\sigma^2_{C_1}/2} \eta_p \epsilon 
- 
\mathrm{e}^{3\mu_{C_3}+9\sigma^2_{C_3}/2} \left(\eta_p \epsilon \right)^3 
+ 
\mathrm{e}^{5\mu_{C_5}+25\sigma^2_{C_5}/2}  \left(\eta_p \epsilon \right)^5
\label{Eqn-Master-StressStrain3}
\end{equation}

\begin{figure}[t]
	\centering
	\includegraphics[width=0.55\textwidth]{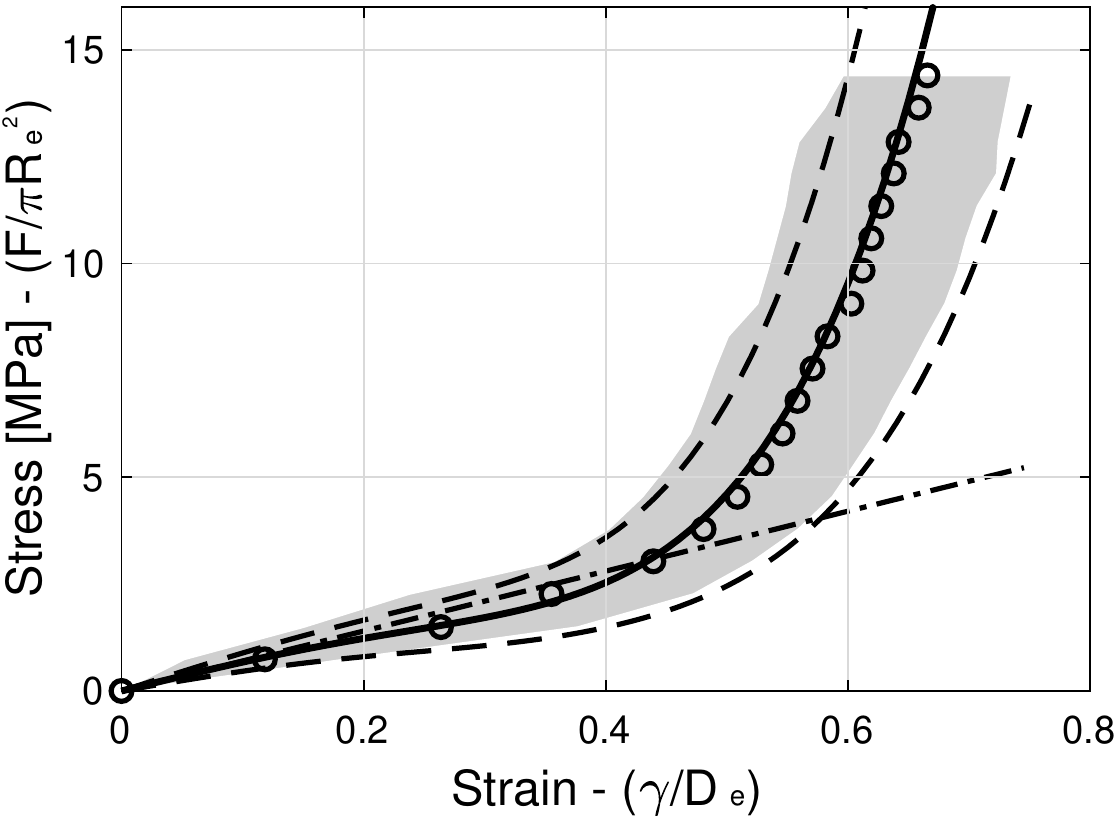}
	\caption{Master contact law $\sigma(\epsilon)$ for micro-crystalline cellulose particles given by Eqn.~\eqref{Eqn-Master-StressStrain2} with coefficients in Table~\ref{Table-StressStrain-Param}. The solid line corresponds to the mean of $\sigma(\epsilon)$ and the dashed lines correspond to one standard deviation from the mean. The dashed-dotted line corresponds to the mean of $\sigma(\epsilon)$ calculated by the first term of the master law. Symbols correspond to the sample mean $\bar{\epsilon}$ and the shaded area corresponds to one sample standard deviation $s_\epsilon$ from the mean. The shape factor $\eta_p$ is given by Eqn.~\eqref{Eqn-ShapeFactor} with $a=1.43$.}
	\label{Fig-MasterPlot-2}
\end{figure}

The approximated stress-strain relationship is readily available by calibration and it is shown in Fig.~\ref{Fig-MasterPlot-2} with coefficients $C_i$ given in Table~\ref{Table-StressStrain-Param}. The expected value is thus given by
\begin{equation}
\mathrm{E}\langle\sigma(\epsilon)\rangle / \mathrm{MPa} 
= 
7.0094~\eta_p \epsilon 
- 
27.235~\left(\eta_p \epsilon \right)^3 
+ 
144.23~\left(\eta_p \epsilon \right)^5
\label{Eqn-Master-StressStrain4}
\end{equation}

The stress-strain relationship given by Eqn.~\eqref{Eqn-Master-StressStrain2} is in spirit of the curvature-corrected nonlocal contact formulation for elastic spherical particles \citep{AGARWAL201826}. The first term in the relationship corresponds to the initial linear deformation behavior (represented by a dashed-dotted curve in Fig.~\ref{Fig-MasterPlot-2}), which has been previously described with contact models developed for spherical indentation \citep{Tabor-1951,Johnson1985,Biwa-1995} and contact of inelastic solids of revolution \citep{Storakers-1997} in the context of small deformations. Of particular interest is the similarity solution by \citet{Biwa-1995}, which describes the stress-strain relationship at the contact of rigid-plastic power law hardening solids. For spherical particles under diametrical compression, the relationship is given by
\begin{equation}
\sigma(\epsilon) = 4(3^{1-1/m})(c^2_{max})^{1+1/2m}\kappa\epsilon^{1+1/2m}
\end{equation}
where $m$ is the power law hardening exponent, $\kappa$ is a representative strength given by \citep{Mesarovic-1999,Mesarovic-2000}
\begin{equation}
\kappa = \sigma^{1-1/m}_yE^{1/m}
\end{equation}
where $\sigma_y$ is the uniaxial yield stress and $E$ is the Young's modulus, and $c^2_{max}$ is a calibrated material parameter, well approximated by the following relationship \citep{Storakers-1994}
\begin{equation}
c^2_{max} = 1.43\exp\left(\frac{-0.97}{m}\right)
\end{equation}
Assuming perfectly plastic material behavior (i.e., $m\to\infty$), the stress-strain relationship reduces to
\begin{equation}
\sigma(\epsilon) = 17.16\sigma_y\epsilon
\label{Eqn-IdealPlasticStorakers}
\end{equation}
and, by correlating the calibrated value of plastic coefficient $C_1$ with the above equation, the uniaxial yield stress is estimated as $\sigma_y =  0.4085\,\mathrm{MPa}$. 
Other terms in the master contact law stress-strain relationship describe moderate-to-large deformation behavior of the particle. Following the initial linear response, a softening or reduction in stiffness is observed as a dip in the slope of the stress-strain curve. This regime of deformations has been termed as the contact interaction regime \citep{Tsigginos2015}, which starts with coalescence of two or more plastically deforming zones surrounding the contacts \citep{Frenning2013,Tsigginos2015}. With the onset of this regime, contacts can no longer be assumed independent. The second term of the master stress-strain relationship describes the contact response in this regime. The third term describes the large deformation strain-hardening response, referred to as geometric hardening \citep{Sundstrom-1973} or the low compressibility regime \citep{Tsigginos2015}, during which the average contact pressure unboundedly rises due to increasing contact interactions. 

\section{Loading-unloading contact law for micro-crystalline cellulose particles}
\label{Section-Loading-Unloading}

Unloading contact laws for elasto-plastic spheres with bonding strength, or adhesion, have been developed \citep{Mesarovic-1999,Mesarovic-2000,Mesarovic-2000b}, assuming elastic perfectly-plastic behavior and using a rigid punch decomposition \citep{Hill-1990}. \citet{Olsson-2013} have extended these laws to elasto-plastic spheres that exhibit power-law plastic hardening behavior, and have verified their validity with detailed finite element simulations. This formulation assumes elastic behavior, approximated by Hooke's law, and Irwin's fracture mechanics to describe elastic recovery of the deformed spheres and the breakage of solid bridges. \citet{Gonzalez2018generalized} developed generalized loading-unloading contact laws for elasto-plastic spheres with bonding strength, which are continuous at the onset of unloading by means of a regularization term, in the spirit of a cohesive zone model. These contact laws are updated incrementally to account for strain path dependency and have been show{n to be numerically robust, efficient, and mechanistically sound in three-dimensional particle mechanics static calculations. Here we endow the master contact law for micro-crystalline cellulose particles, proposed in Section~\ref{Section-MasterLaw}, with elastic relaxation during unloading. Specifically, we follow \citet{Gonzalez2018generalized} and modify Eqn.~\eqref{Eqn-Master-StressStrain2} accordingly, i.e.
\begin{equation}
\sigma(\epsilon)
=
\left\{
\begin{array}{l}
\left[ C_1~\eta_p \epsilon - \left( C_3~\eta_p \epsilon \right)^3 + \left( C_5~\eta_p \epsilon \right)^5 \right]\mbox{MPa}
\hspace{0.90in}
\mbox{plastic loading}
\\
\frac{2}{\pi} \sigma_{\mbox{\tiny P}}
\left[
\arcsin\left( \phi(\epsilon, \epsilon_{\mbox{\tiny P}} ) \right)
-
\phi(\epsilon, \epsilon_{\mbox{\tiny P}})  \sqrt{1-\phi(\epsilon, \epsilon_{\mbox{\tiny P}})^2}
\right]
\hspace{0.35in}
\mbox{elastic (un)loading}
\end{array}
\right.
\label{Eqn-MasterLoadingUnloading}
\end{equation}
where the internal variables are updated, i.e., $\{\epsilon_{\mbox{\tiny P}},\sigma_{\mbox{\tiny P}} \} \leftarrow  \{\epsilon,\sigma(\epsilon)\} $, during plastic loading (i.e., when $\epsilon \ge \epsilon_{\mbox{\tiny P}}$). In the above equation, the function $\phi(\epsilon, \epsilon_{\mbox{\tiny P}})$ is given by
\begin{equation}
\phi(\epsilon, \epsilon_{\mbox{\tiny P}})
=
\left[1 - 4.12 \bar{E}  \frac{(\epsilon_{\mbox{\tiny P}} - \epsilon)^2 \epsilon_{\mbox{\tiny P}}}{\sigma_{\mbox{\tiny P}}}  \right]_+^{1/2}
\end{equation}
for an elastic, perfectly plastic particle (i.e., for $m\rightarrow\infty$) with effective elastic modulus equal to $\bar{E}=E/(1-\nu^2)=16$~GPa. The assumption of perfectly plastic material is in agreement with the observation made in Section~\ref{Section-MasterLaw} and the value of the effective elastic modulus is the result of a simple calibration.

\section{Results and discussion}
\label{Section-Results}

We have proposed that the experimentally characterized micro-crystalline cellulose particles deform under diametrical compression following a master contact law $\sigma(\epsilon)$, given by Eqn.~\eqref{Eqn-MasterLoadingUnloading} with plastic coefficients $C_i$ in Table~\ref{Table-StressStrain-Param}, effective elastic modulus $\bar{E}=16$~GPa, and a shape factor $\eta_p$, given by Eqn.~\eqref{Eqn-ShapeFactor} with $a=1.43$. Figs.~\ref{Fig-ExpVSModel-1}, \ref{Fig-ExpVSModel-2} and \ref{Fig-ExpVSModel-3} exhibit a very good agreement between predictions of the calibrated contact law and the experimental values. It is evident from the figures that the apparent plastic strain-hardening occurs at distinctly different strain values and that the proposed contact law captures its dependency on particle dimensions---i.e., on $D_1$, $D_2$ and $D_3$ (see Fig.~\ref{Fig-DimensionStatistics}). It is also evident that the elastic relaxation during unloading is accurately predicted with low uncertainty. 

\section{Summary}
\label{Section-Summary}

We have proposed a semi-empirical mechanistic contact law for micro-crystalline cellulose (Avicel PH-200) particles. This loading-unloading contact law has been characterized experimentally using diametrical compression force-displacement curves, obtained with a Shimadzu MCT-510 micro-compression tester. The irregular MCC particles have been approximated by an ellipsoid and the lengths of their principal axes have been measured using an in-built microscope and a side camera. To generalize the contact mechanics of an elastic ellipsoidal particle to an elasto-plastic irregular particle approximated by an ellipsoid, we have introduced the concepts of a shape factor and a master contact law. It is worth noting that the force-displacement curves exhibit an apparent strain-hardening at distinctly different strain values. We have postulated that this deformation mechanism depends on particle dimensions and loading configuration,which is captured by the shape factor function. The proposed loading-unloading contact law is, therefore, a function of (i) three characteristic diameters (lengths of the principal axes of an approximated ellipsoid), (ii) a geometric parameter associated with the loading condition through the shape factor function, (iii) three plastic material properties, and (iv) one effective elastic material property. The three plastic material properties are log-normal distributions estimated from the loading experimental curves, while the effective elastic property is estimated from the unloading experimental curves. It bears emphasis that the proposed loading contact law is in spirit of the curvature-corrected nonlocal contact formulation for elastic spherical particles \citep{AGARWAL201826}, with the first term correponding to the stress-strain relationship described by small-strain contact models developed for spherical indentation \citep{Tabor-1951,Johnson1985,Biwa-1995} and contact of inelastic solids of revolution \citep{Storakers-1997}. Similarly, the proposed unloading contact law follows from generalized loading-unloading contact laws for elasto-plastic spheres with bonding strength \citep{Gonzalez2018generalized}. The study shows a very good agreement between predictions of the calibrated loading-unloading contact law and the experimental values. However, it is recommended that a larger number of experiments be performed to estimate the five parameters of the contact law, in order to reduce the uncertainty in the model predictions. 

We close by pointing out that the proposed semi-empirical mechanistic contact law is relevant to three-dimensional particle mechanics calculations \citep{Gonzalez2012,gonzalez2016microstructure,yohannes2016evolution,yohannes2017discrete,gonzalez2018statistical,Gonzalez2018generalized}. Therefore, the work presented in this paper, in combination with these detailed calculations, can contribute to develop microstructure-mediated process-structure-property-performance interrelationships and, thus, to establish the relationship between particle-level material properties and tablet performance. Ultimately, these relationships are needed to assist QbD and QbC product development and process control \citep{Yi2018}.
	
\begin{figure}[p!]
	\centering
	\begin{subfigure}[b]{0.32\textwidth}
		\captionsetup{justification=centering}
		\includegraphics[width=\linewidth]{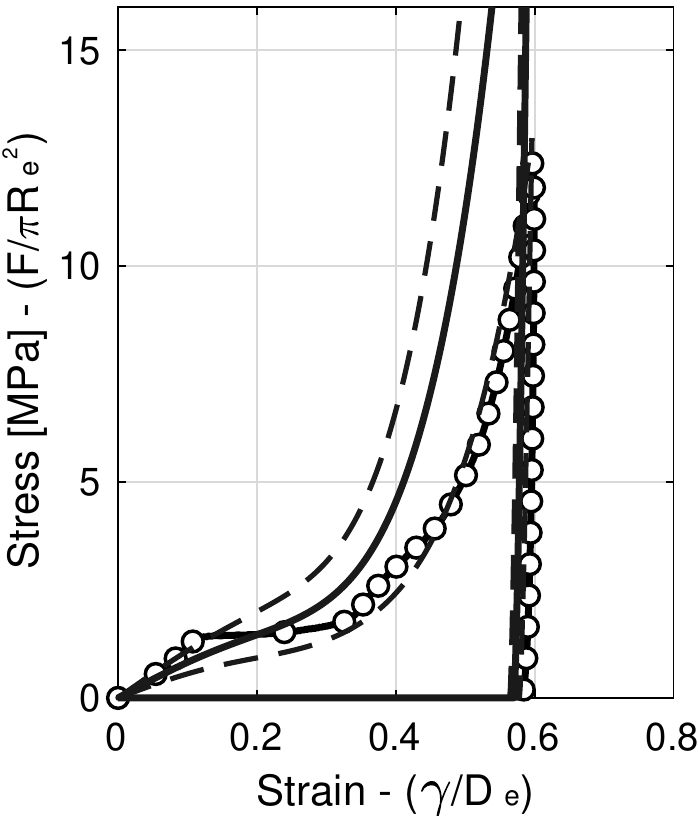}
		\caption{Particle 1}
		\label{fig:Fig-1}
	\end{subfigure}
	\hfill
	\begin{subfigure}[b]{0.32\textwidth}
		\captionsetup{justification=centering}
		\includegraphics[width=\linewidth]{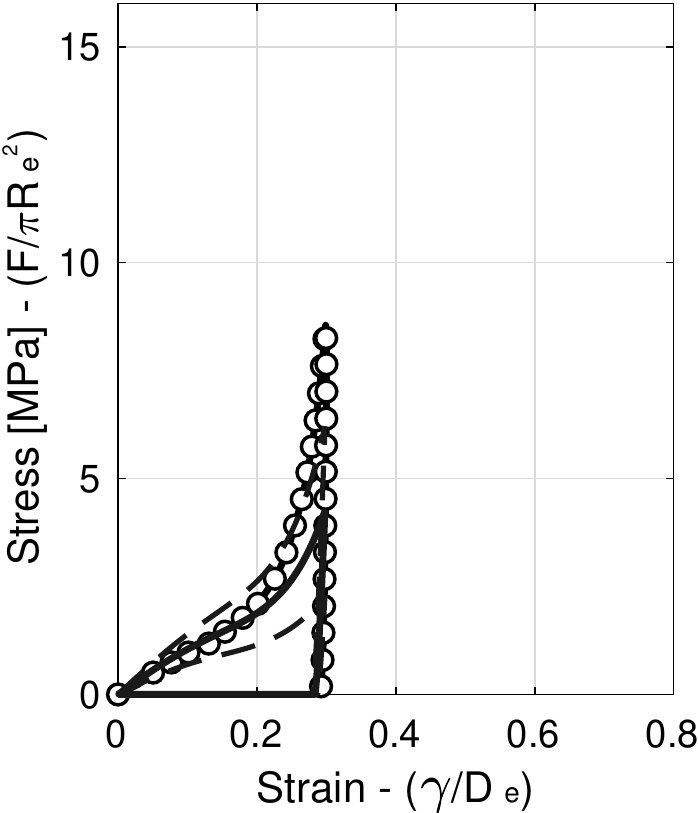}
		\caption{Particle 2}
		\label{fig:Fig-2}
	\end{subfigure}
	\hfill
	\begin{subfigure}[b]{0.32\textwidth}
		\captionsetup{justification=centering}
		\includegraphics[width=\linewidth]{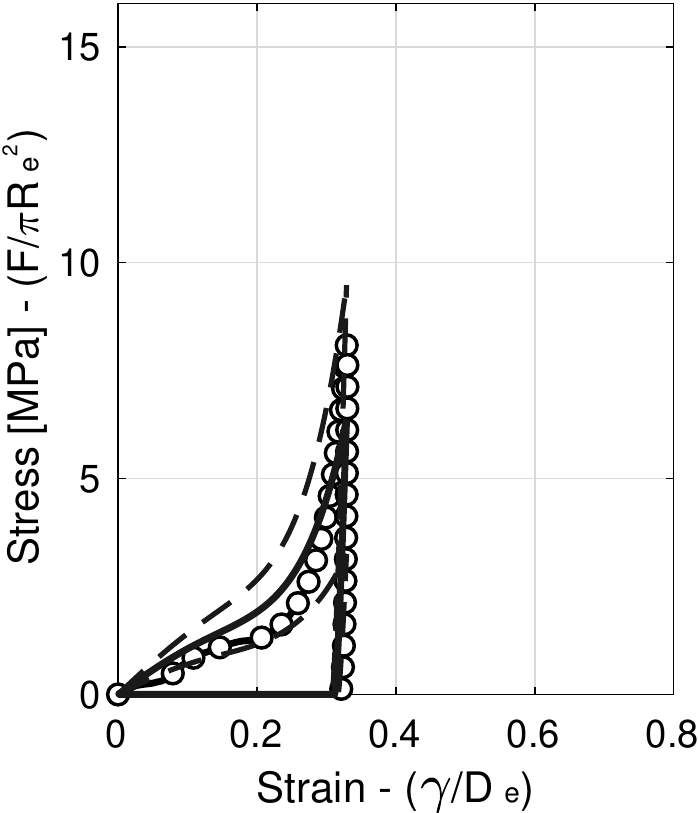}
		\caption{Particle 3}
		\label{fig:Fig-3}
	\end{subfigure}
	\par\bigskip
	\begin{subfigure}[b]{0.32\textwidth}
		\captionsetup{justification=centering}
		\includegraphics[width=\linewidth]{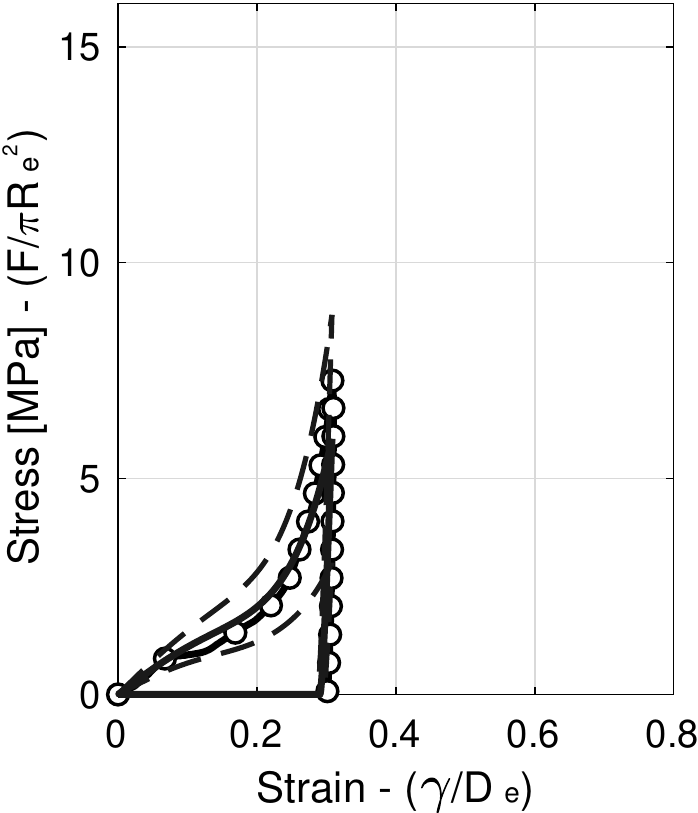}
		\caption{Particle 4}
		\label{fig:Fig-4}
	\end{subfigure}
	\hfill
	\begin{subfigure}[b]{0.32\textwidth}
		\captionsetup{justification=centering}
		\includegraphics[width=\linewidth]{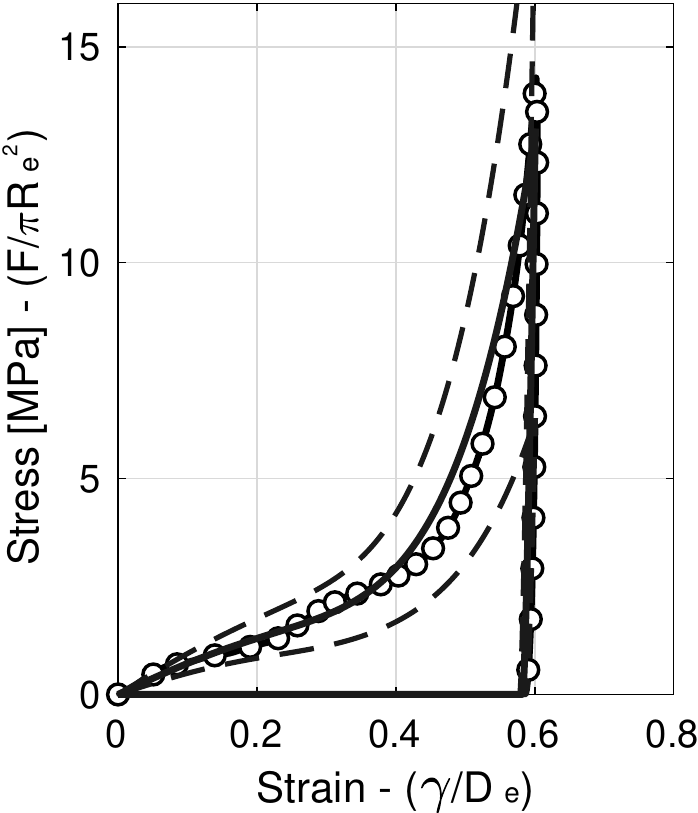}
		\caption{Particle 5}
		\label{fig:Fig-5}
	\end{subfigure}
	\hfill
	\begin{subfigure}[b]{0.32\textwidth}
		\captionsetup{justification=centering}
		\includegraphics[width=\linewidth]{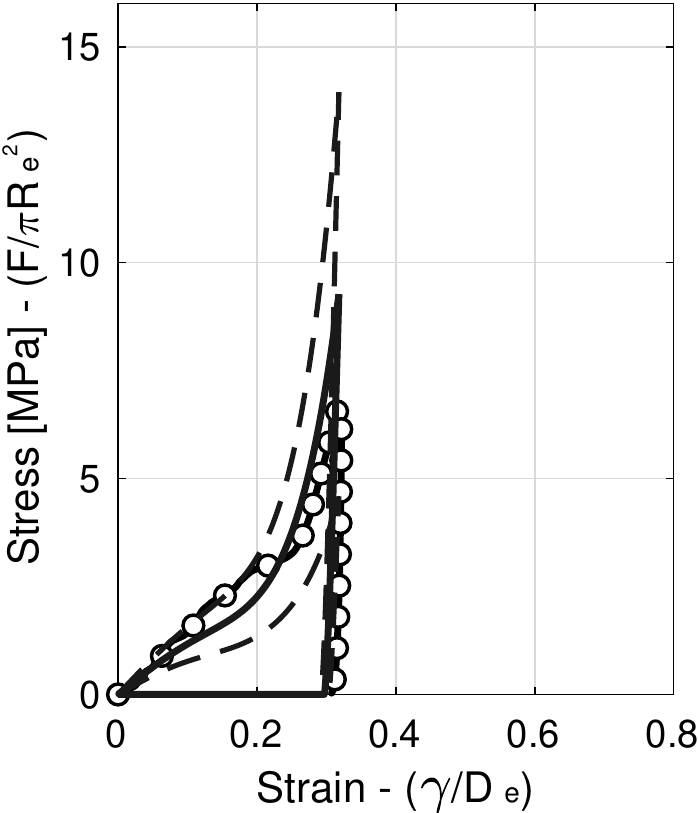}
		\caption{Particle 6}
		\label{fig:Fig-6}
	\end{subfigure}
	\par\bigskip
	\begin{subfigure}[b]{0.32\textwidth}
		\captionsetup{justification=centering}
		\includegraphics[width=\linewidth]{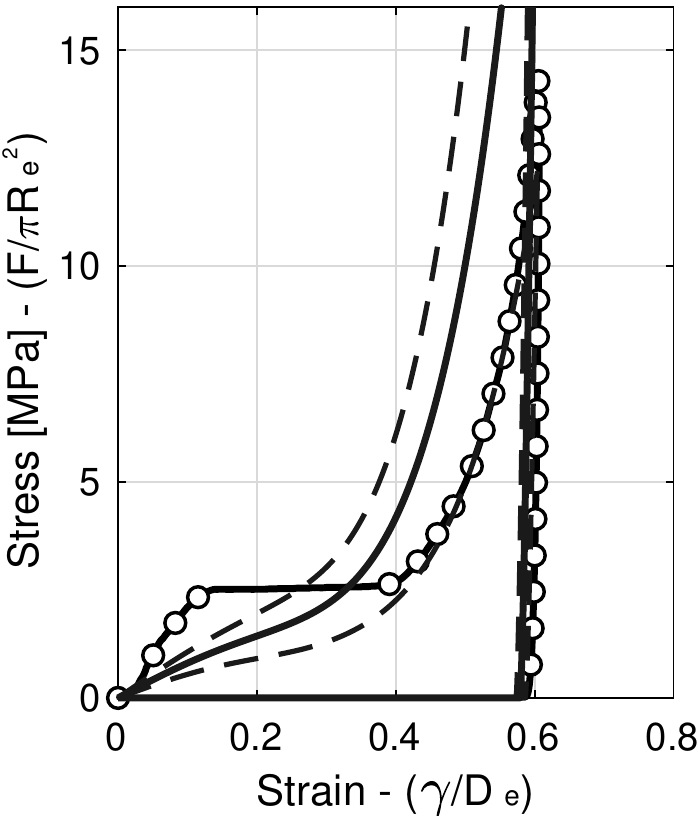}
		\caption{Particle 7}
		\label{fig:Fig-7}
	\end{subfigure}
	\hfill
	\begin{subfigure}[b]{0.32\textwidth}
		\captionsetup{justification=centering}
		\includegraphics[width=\linewidth]{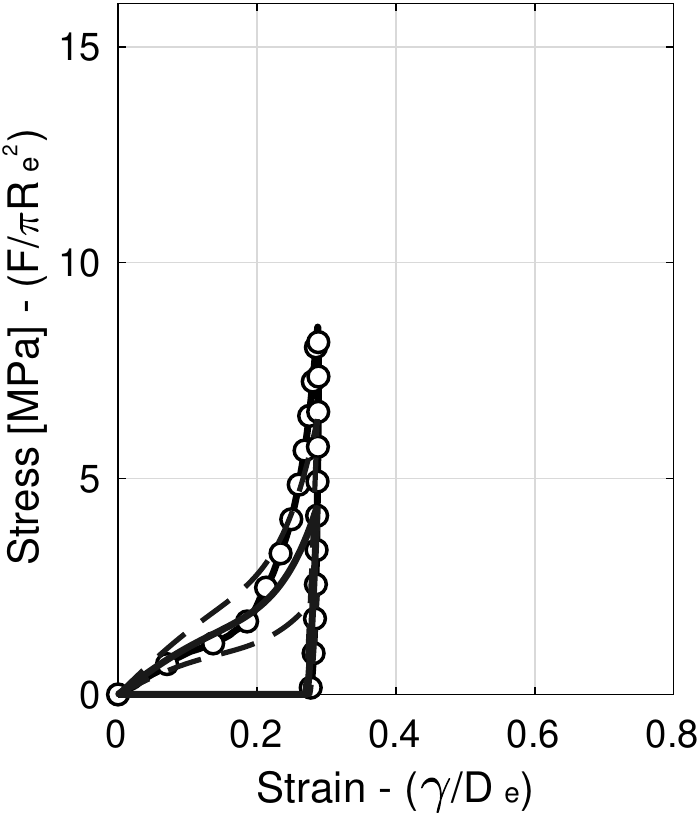}
		\caption{Particle 8}
		\label{fig:Fig-8}
	\end{subfigure}
	\hfill
	\begin{subfigure}[b]{0.32\textwidth}
		\captionsetup{justification=centering}
		\includegraphics[width=\linewidth]{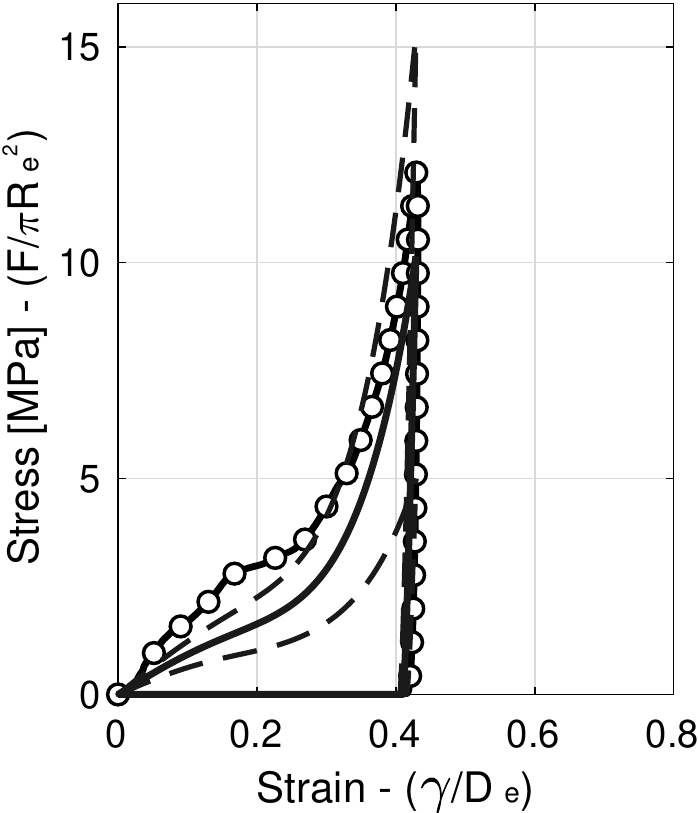}
		\caption{Particle 9}
		\label{fig:Fig-9}
	\end{subfigure}
	\caption{Experimental values and predictions of loading-unloading contact curves for micro-crystalline cellulose (Avicel PH-200) particles under diametrical compression.  The master contact law $\sigma(\epsilon)$  is given by Eqn.~\eqref{Eqn-MasterLoadingUnloading} with plastic coefficients $C_i$ in Table~\ref{Table-StressStrain-Param}, effective elastic modulus $\bar{E}=16$~GPa, and the shape factor $\eta_p$ is given by Eqn.~\eqref{Eqn-ShapeFactor} with $a=1.43$.}
	\label{Fig-ExpVSModel-1}
\end{figure} 

\begin{figure}[p!]
	\centering
	\begin{subfigure}[b]{0.32\textwidth}
		\captionsetup{justification=centering}
		\includegraphics[width=\linewidth]{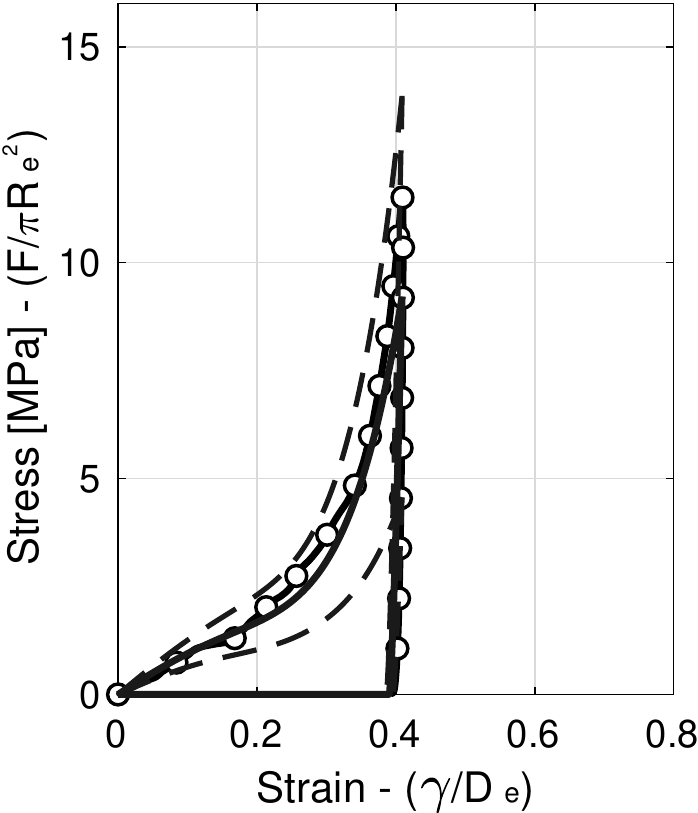}
		\caption{Particle 10}
		\label{fig:Fig-10}
	\end{subfigure}
	\hfill
	\begin{subfigure}[b]{0.32\textwidth}
		\captionsetup{justification=centering}
		\includegraphics[width=\linewidth]{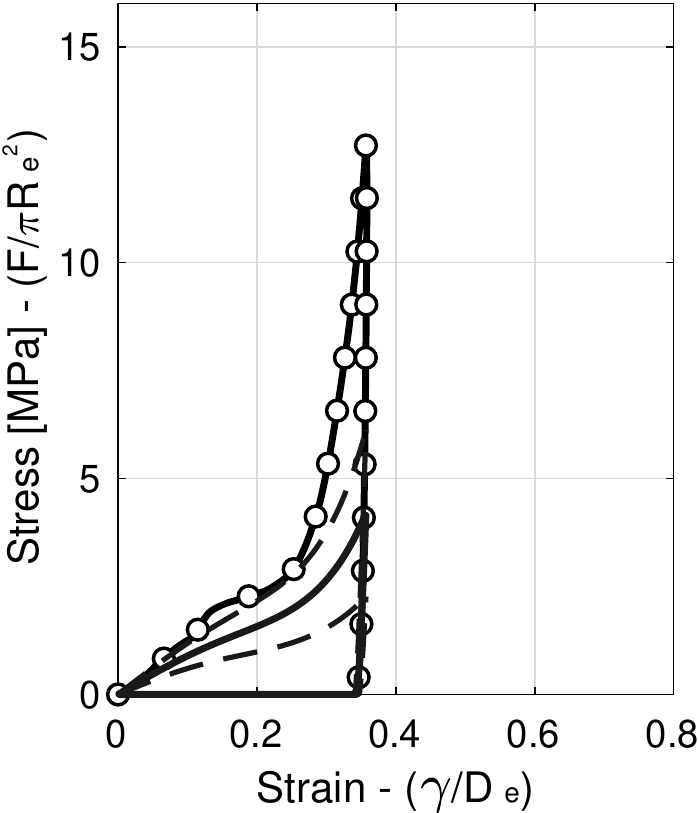}
		\caption{Particle 11}
		\label{fig:Fig-11}
	\end{subfigure}
	\hfill
	\begin{subfigure}[b]{0.32\textwidth}
		\captionsetup{justification=centering}
		\includegraphics[width=\linewidth]{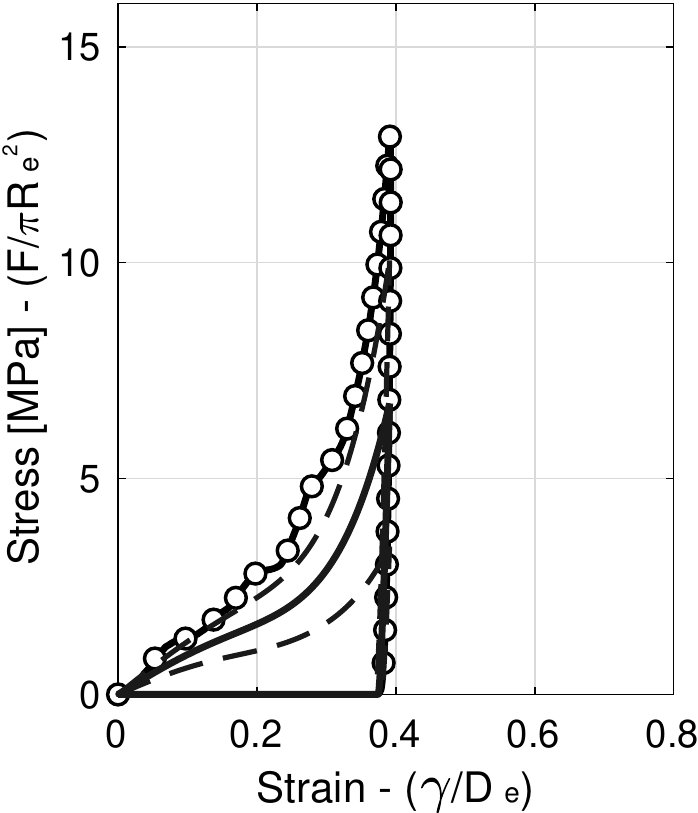}
		\caption{Particle 12}
		\label{fig:Fig-12}
	\end{subfigure}
	\par\bigskip
	\begin{subfigure}[b]{0.32\textwidth}
		\captionsetup{justification=centering}
		\includegraphics[width=\linewidth]{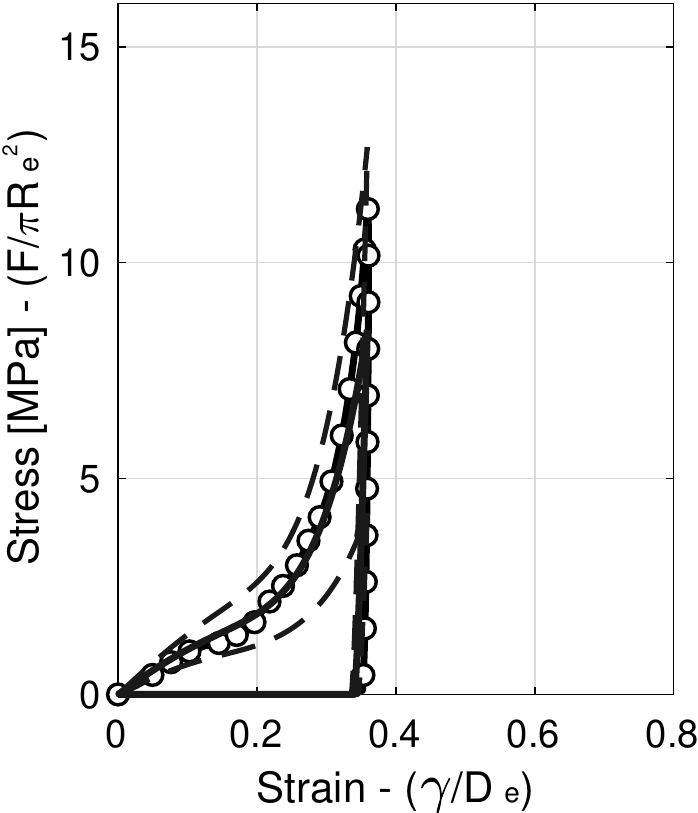}
		\caption{Particle 13}
		\label{fig:Fig-13}
	\end{subfigure}
	\hfill
	\begin{subfigure}[b]{0.32\textwidth}
		\captionsetup{justification=centering}
		\includegraphics[width=\linewidth]{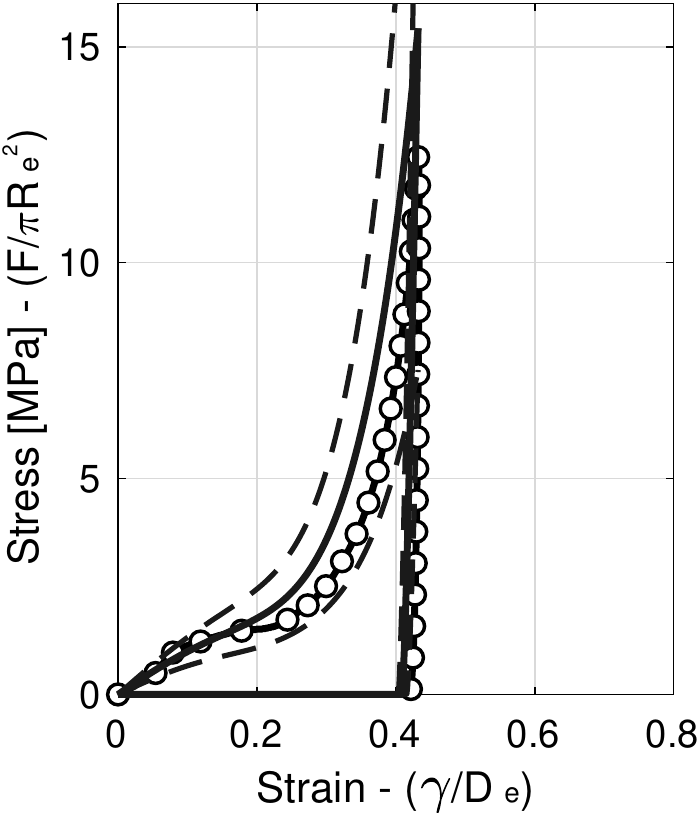}
		\caption{Particle 14}
		\label{fig:Fig-14}
	\end{subfigure}
	\hfill
	\begin{subfigure}[b]{0.32\textwidth}
		\captionsetup{justification=centering}
		\includegraphics[width=\linewidth]{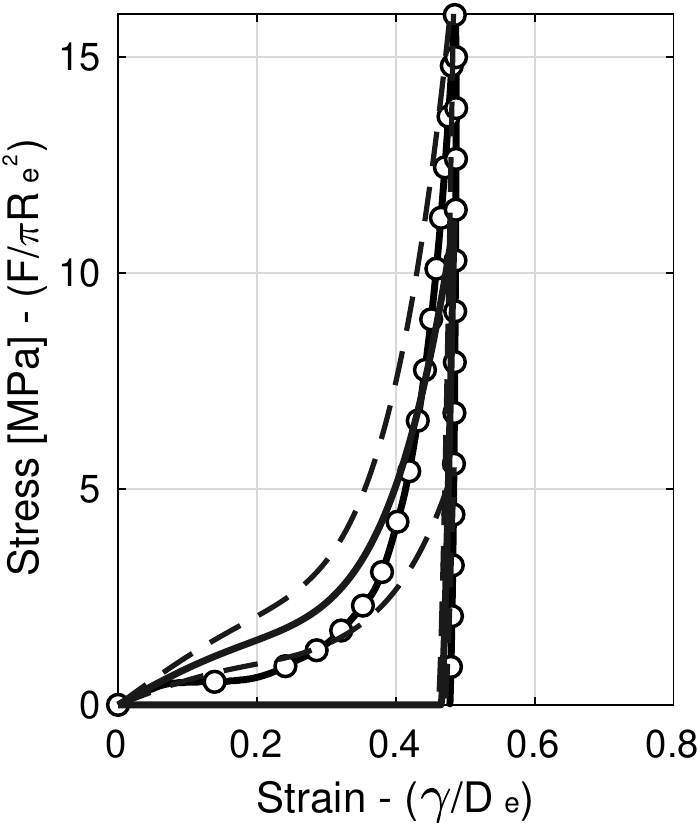}
		\caption{Particle 15}
		\label{fig:Fig-15}
	\end{subfigure}
	\par\bigskip
	\begin{subfigure}[b]{0.32\textwidth}
		\captionsetup{justification=centering}
		\includegraphics[width=\linewidth]{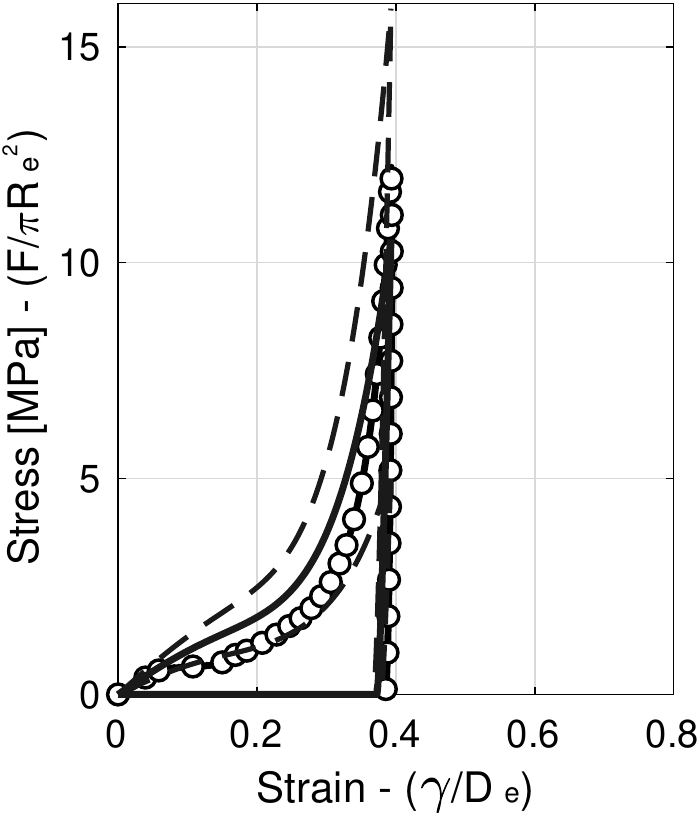}
		\caption{Particle 16}
		\label{fig:Fig-16}
	\end{subfigure}
	\hfill
	\begin{subfigure}[b]{0.32\textwidth}
		\captionsetup{justification=centering}
		\includegraphics[width=\linewidth]{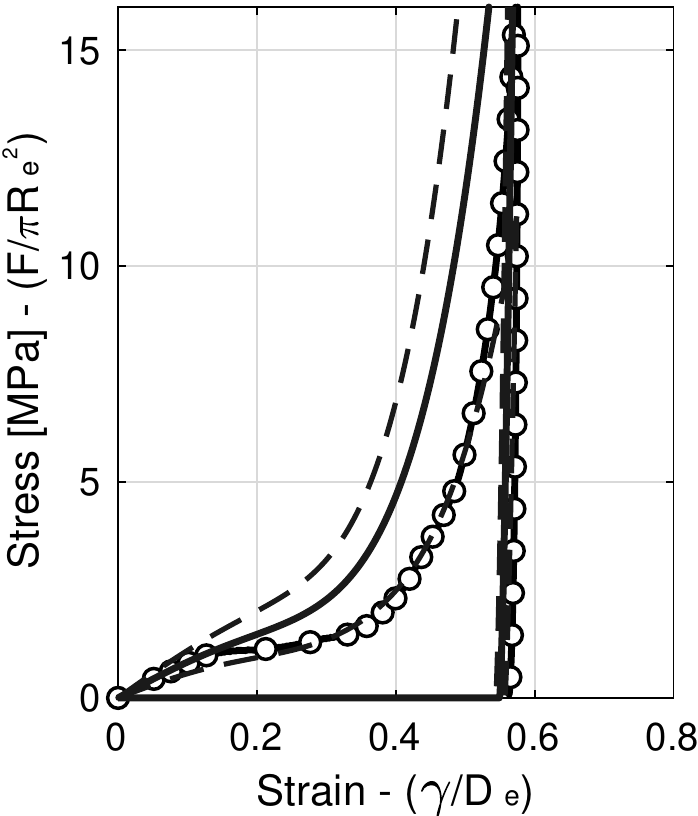}
		\caption{Particle 17}
		\label{fig:Fig-17}
	\end{subfigure}
	\hfill
	\begin{subfigure}[b]{0.32\textwidth}
		\captionsetup{justification=centering}
		\includegraphics[width=\linewidth]{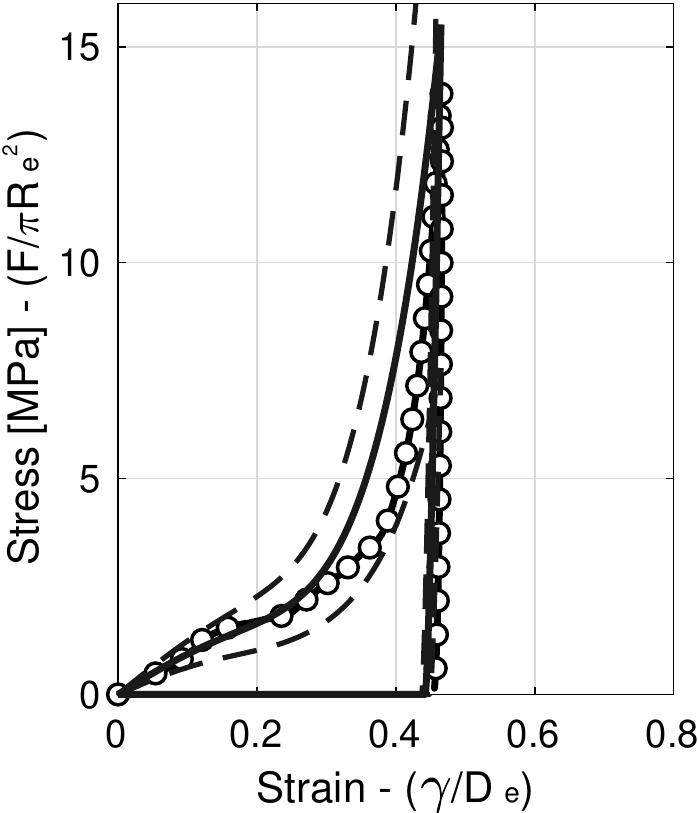}
		\caption{Particle 18}
		\label{fig:Fig-18}
	\end{subfigure}
	\caption{Experimental values and predictions of loading-unloading contact curves for micro-crystalline cellulose (Avicel PH-200) particles under diametrical compression.  The master contact law $\sigma(\epsilon)$  is given by Eqn.~\eqref{Eqn-MasterLoadingUnloading} with plastic coefficients $C_i$ in Table~\ref{Table-StressStrain-Param}, effective elastic modulus $\bar{E}=16$~GPa, and the shape factor $\eta_p$ is given by Eqn.~\eqref{Eqn-ShapeFactor} with $a=1.43$.}
	\label{Fig-ExpVSModel-2}
\end{figure}

\begin{figure}[t]
	\centering
	\begin{subfigure}[b]{0.32\textwidth}
		\captionsetup{justification=centering}
		\includegraphics[width=\linewidth]{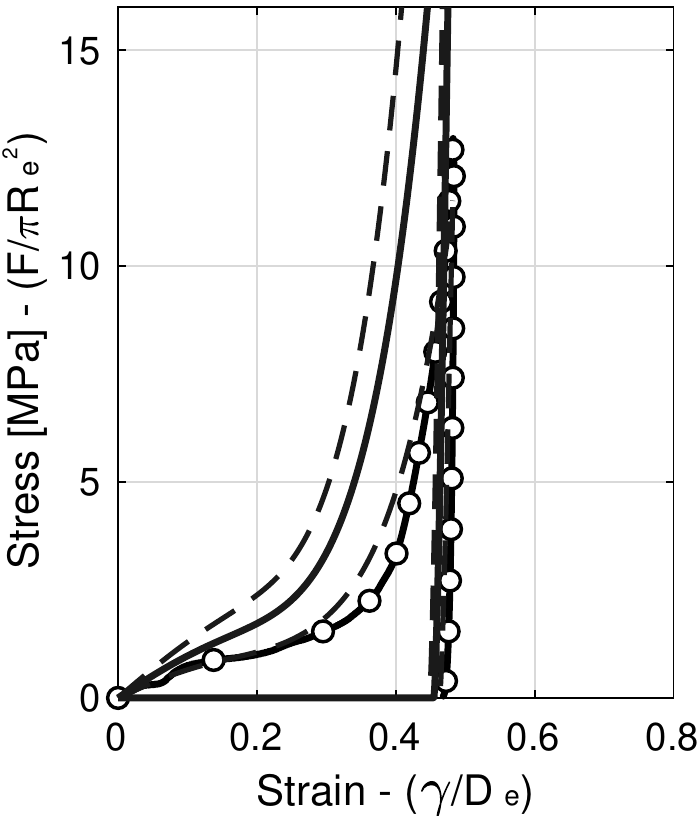}
		\caption{Particle 19}
		\label{fig:Fig-19}
	\end{subfigure}
	\hspace{50pt}
	\begin{subfigure}[b]{0.32\textwidth}
		\captionsetup{justification=centering}
		\includegraphics[width=\linewidth]{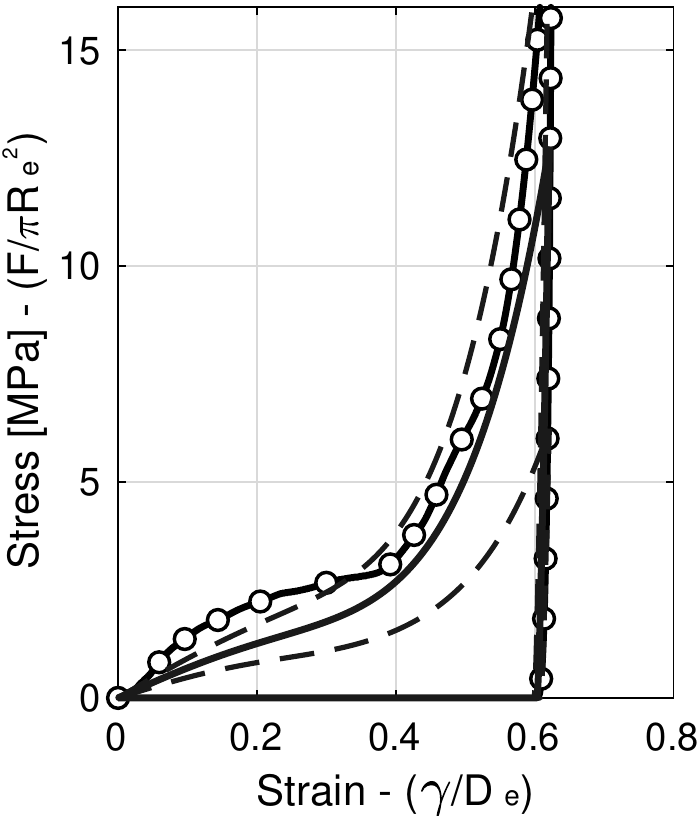}
		\caption{Particle 20}
		\label{fig:Fig-20}
	\end{subfigure}
	\caption{Experimental values and predictions of loading-unloading contact curves for micro-crystalline cellulose (Avicel PH-200) particles under diametrical compression.  The master contact law $\sigma(\epsilon)$  is given by Eqn.~\eqref{Eqn-MasterLoadingUnloading} with plastic coefficients $C_i$ in Table~\ref{Table-StressStrain-Param}, effective elastic modulus $\bar{E}=16$~GPa, and the shape factor $\eta_p$ is given by Eqn.~\eqref{Eqn-ShapeFactor} with $a=1.43$.}
	\label{Fig-ExpVSModel-3}
\end{figure}

\section*{Acknowledgements}

The author gratefully acknowledges the support received from the United States National Science Foundation grant number CMMI-1538861 and from the United States Food and Drug Administration grant number DHHS-FDA U01FD005535. The views expressed by authors do not necessarily reflect the official policies of the Department of Health and Human Services; nor does any mention of trade names, commercial practices, or organization imply endorsement by the United States Government.

\section*{Nomenclature}

\noindent
\begin{longtable}{ll}	
	$a$ & geometric factor for plastic ellipsoidal particles (-)\\
	$c^2_{max}$ & material parameter for similarity contact law (-)\\
	$C_1$ & plastic material property 1 ($\mathrm{MPa}$) \\
	$C_2$ & plastic material property 2 ($\mathrm{MPa}^{-3}$)  \\
	$C_3$ & plastic material property 2 ($\mathrm{MPa}^{-5}$)  \\
	$D_1$ & major diameter of the ellipsoidal particle when looked through MCT's microscope ($\mathrm{mm}$)\\
	$D_2$ & minor diameter of the ellipsoidal particle when looked through MCT's microscope ($\mathrm{mm}$)\\
	$D_3$ & vertical diameter of the ellipsoidal particle captured by MCT's side-view camera ($\mathrm{mm}$)\\
	$E$ & Young's modulus ($\mathrm{MPa}$)\\
	$\mathrm{E}[\cdot]$ & expected value (-)\\
	$F$ & force acting on the particle ($\mathrm{N}$)\\
	$R_e$ & effective radius of the ellipsoidal particle ($\mathrm{mm}$)\\
	$m$ & Power-law hardening exponent (-)\\
	$N$ & total number of tested particles (-)\\
	$s_\epsilon$ & standard deviation of a set of strain values (-)\\
	$\#\mathcal{S}_\sigma$ & number of experiments for which a given stress $\sigma$ is reached (-)\\
	$\gamma$ & particle deformation at the contact ($\mathrm{mm}$)\\
	$\epsilon$ & effective particle strain (-)\\	
	$\bar{\epsilon}$ & mean of the strain values for a set of particles (-)\\
	$\epsilon_p$ & internal strain variable for unloading contact law (-)\\
	$\eta_e$ & elastic shape factor (-)\\
	$\eta_p$ & plastic shape factor (-)\\	
	$\kappa$ & particle reference strength ($\mathrm{MPa}$) \\
	$\nu$ & Poisson's ratio (-)\\	
	$\sigma$ & effective particle stress ($\mathrm{MPa}$)\\
	$\sigma^m$ & maximum stress value in the master contact law ($\mathrm{MPa}$)\\
	$\sigma_y$ & uniaxial yield stress ($\mathrm{MPa}$)\\
	$\sigma_p$ & internal stress variable for unloading contact law ($\mathrm{MPa}$) \\	
\end{longtable}

\bibliographystyle{kona}
\bibliography{all}

\end{document}